\documentclass{article}

\usepackage{amssymb}
\usepackage{latexsym}

\usepackage{url}
\usepackage[dvipsnames]{xcolor}
\definecolor{newcolor}{rgb}{.8,.349,.1}
\usepackage{multirow}
\usepackage{color}
\usepackage{float}
\usepackage[utf8]{inputenc}  %for french characters
\usepackage{subfiles} %to separate tex to subfiles
\usepackage[square,numbers,sort&compress]{natbib}
\usepackage{amssymb}
\usepackage{mathtools}
\usepackage{gensymb}
\usepackage{authblk}
\usepackage[margin=0.8in]{geometry}
\usepackage[font=footnotesize]{subfig}
\usepackage{subfiles}
\usepackage{amsmath}
\usepackage[pdftex]{graphicx}

\usepackage{array}
\newcolumntype{M}[1]{>{\centering\arraybackslash}m{#1}}

\usepackage{hyperref}

\usepackage{multicol}
\usepackage{longtable}
\usepackage{forest}
\usepackage{adjustbox}
\usepackage{diagbox}
\usepackage{tikz}
\usetikzlibrary{arrows.meta, shapes.geometric, calc, shadows}
\usetikzlibrary{shapes,arrows,shadows, snakes}
\usetikzlibrary{shapes.arrows}

\colorlet{mygreen}{green!75!black}
\colorlet{col1in}{red!30}
\colorlet{col1out}{red!40}
\colorlet{col2in}{mygreen!30}
\colorlet{col2out}{mygreen!40}
\colorlet{col2}{mygreen}
\colorlet{col3in}{blue!30}
\colorlet{col3out}{blue!40}
\colorlet{col3}{blue}
\colorlet{col4in}{mygreen!20}
\colorlet{col4out}{mygreen!30}
\colorlet{col5in}{blue!10}
\colorlet{col5out}{blue!20}
\colorlet{col6in}{blue!20}
\colorlet{col6out}{blue!30}
\colorlet{col7out}{orange}
\colorlet{col7in}{orange!50}
\colorlet{col8out}{orange!40}
\colorlet{col8in}{orange!20}
\colorlet{linecol}{blue!60}

\pgfkeys{/forest,
  rect/.append style   = {rectangle, rounded corners = 2pt,
                         inner color = col6in, outer color = col6out},
  rectpy/.append style   = {rectangle, rounded corners = 2pt,
                         inner color = col2in, outer color = col2in},
  ellip/.append style  = {ellipse, inner color = col5in,
                          outer color = col5out},
  orect/.append style  = {rect, font = \sffamily\bfseries\LARGE,
                         text width = 325pt, text centered,
                         minimum height = 10pt, outer color = col7out,
                         inner color=col7in},
  oellip/.append style = {ellip, inner color = col8in, outer color = col8out,
                          font = \sffamily\bfseries\large, text centered}}
\usepackage{verbatim}
\usepackage{smartdiagram}
\usepackage{metalogo}
\usepackage{float}
\usepackage{listings}

\definecolor{backcolour}{rgb}{0.95,0.95,0.92}

\lstdefinestyle{mystyle}{
    backgroundcolor=\color{white}, 
    breakatwhitespace=false,         
    breaklines=true,                 
    captionpos=b,                    
    keepspaces=true,                 
    showtabs=false,                  
    tabsize=2
}
\RequirePackage{subfig}
\RequirePackage{xspace}

\newcommand{\telemacsystem}{{\scshape te\-le\-mac-ma\-sca\-ret sy\-st\-em}\xspace}
\newcommand{\telemacmascaret}{{\scshape te\-le\-mac-ma\-sca\-ret}\xspace}
\newcommand{\telemacdd}{{\scshape te\-le\-mac-2d}\xspace}
\newcommand{\telemacddd}{{\scshape te\-le\-mac-3d}\xspace}
\newcommand{\gaia}{{\scshape gaia}\xspace}
\newcommand{\courlis}{{\scshape cou\-rl\-is}\xspace}
\newcommand{\tracer}{{\scshape tra\-cer}\xspace}
\newcommand{\tomawac}{{\scshape to\-ma\-wac}\xspace}
\newcommand{\artemis}{{\scshape ar\-tem\-is}\xspace}
\newcommand{\waqtel}{{\scshape waq\-tel}\xspace}
\newcommand{\mascaret}{{\scshape ma\-sca\-ret}\xspace}

\newcommand{\hecras}{{\scshape hec-ras}\xspace}
\newcommand{\mohid}{{\scshape mohid}\xspace}
\newcommand{\delft}{{\scshape delft-3d}\xspace}

\lstdefinelanguage{TelFortran}{                                                                          
  language=[90]Fortran,
  basicstyle=\ttfamily\small,
  frameround=ffff,
  stringstyle={\color{magenta}},
  showstringspaces=false,
  morekeywords={USE,PRECISION,INTENT,IN,OUT,INOUT,ONLY,ADVANCE,WHILE,POINTER,MODULE},
  deletekeywords=[1]{REAL},
  keywordstyle=\bfseries,
  commentstyle=\color{gray},
  morecomment=[l]{!},
  morekeywords=[2]{SQRT,REAL,INT,MAX,MIN},
  keywordstyle=[2]{\bfseries\color{violet}},
  morekeywords=[3]{ALLOCATED,PRESENT,DEALLOCATE,ALLOCATE},
  keywordstyle={\color{blue}},
  literate={%
    *{.EQ.}{{{\bfseries\color{green}.EQ.}}}{4}
     {.NE.}{{{\bfseries\color{green}.NE.}}}{4}
     {.LE.}{{{\bfseries\color{green}.LE.}}}{4}
     {.LT.}{{{\bfseries\color{green}.LT.}}}{4}
     {.GE.}{{{\bfseries\color{green}.GE.}}}{4}
     {.GT.}{{{\bfseries\color{green}.GT.}}}{4}
     {.NOT.}{{{\bfseries\color{green}.NOT.}}}{5}
     {.OR.}{{{\bfseries\color{green}.OR.}}}{4}
     {.AND.}{{{\bfseries\color{green}.AND.}}}{5}
     {.TRUE.}{{{\bfseries\color{violet}.TRUE.}}}{6}
     {.FALSE.}{{{\bfseries\color{violet}.FALSE.}}}{7}
  },
    escapeinside={(*@}{@*)},
}

\lstdefinelanguage{TelPython}{                                                                          
  language=Python,
  basicstyle=\ttfamily\small,
  frameround=ffff,
  stringstyle={\color{magenta}},
  showstringspaces=false,
  commentstyle=\color{gray},
  keywordstyle={\color{blue}},
}
\usepackage{tabularx}
\usepackage{soul}
\sethlcolor{green}
\begin{document}

\title{Interoperability and computational framework for simulating open channel hydraulics: application to sensitivity analysis and calibration of Gironde Estuary model}%

\author{Cédric Goeury}
\author{Yoann Audouin}
\author{Fabrice Zaoui}

\affil{\'{E}lectricité de France (EDF), Research and Development Division, National Laboratory for Hydraulics and Environment (LNHE), 6 Quai Watier, 78400 Chatou, France}

\date{\today}
\maketitle

\begin{abstract}
Water resource management is of crucial societal and economic importance, requiring a strong capacity for anticipating environmental change. Progress in physical process knowledge, numerical methods and computational power, allows us to address hydro-environmental problems of growing complexity. Modeling of river and marine flows is no exception. With the increase in IT resources, environmental modeling is evolving to meet the challenges of complex real-world problems. This paper presents a new distributed Application Programming Interface (API) of the open source \telemacsystem to run hydro-environmental simulations with the help of the interoperability concept. Use of the API encourages and facilitates the combination of worldwide reference environmental libraries with the hydro-informatic system. Consequently, the objective of the paper is to promote the interoperability concept for studies dealing with such issues as uncertainty propagation, global sensitivity analysis, optimization, multi-physics or multi-dimensional coupling. To illustrate the capability of the API, an operational problem for improving the navigation capacity of the Gironde Estuary is presented. The API potential is demonstrated in a re-calibration context. The API is used for a multivariate sensitivity analysis to quickly reveal the most influential parameters which can then be optimally calibrated with the help of a data assimilation technique.
\end{abstract}

%% main text
%===========================================================================
\section*{Software availability}

The Application Programing Interface (API) framework described in this article (Fortran APIs and its Python wrapper) is available for download in \telemacsystem (\url{www.opentelemac.org}). \telemacmascaret is an integrated suite of solvers for use in the field of free-surface flow, available under the GNU General Public License version 3.

\section{Introduction}

Water availability and quality are vital to human health. Increasing global scarcity makes anticipating the evolution of this limited natural resource even more essential. This ability relies on the understanding and prediction of hydrodynamic flows. Hydraulic simulation codes (such as \hecras \citep{dyhouse_2007}, \mohid \citep{Mateus_2013}, \delft \citep{Deltares_2014}, \telemacmascaret \citep{Hervouet_2007}) have been developed for many years and applied to a wide range of hydro-environmental cases, such as prevention of flood risks \citep{Teng_2017}, development of management plans \citep{Ran_2016} or design schemes for flood alleviation \citep{Ogie_2020}.

Hydro-environmental modeling of river and marine flows is increasingly complex and requires a better knowledge of physical and biochemical processes, advanced numerical methods and high-performance computing resources. For many applications, engineers have to deal with complex ``modeling systems'' for the representation of subparts of a physical system \cite{Braunschweig_2004,Harpham_2014}. Increasing a numerical model conceptual complexity in terms of how many aspects of the underlying system are included in it has the potential to improve the accuracy of the results in term of descriptive and predictive capacities. However, adding each new aspect to the model can induce an increase of the error brought about by the uncertainty in the description of the features accumulates and increases the overall uncertainty in the output \cite{CEA_2011,Lee_2000} . This trade-off between ``model completeness'' and ``propagation error'' \citep{Saltelli_2019} is known as the conjecture of O’Neill \citep{Oneil_1979}. In maritime or fluvial cases, numerical modeling tools allow replicating the past as well as predicting the future with an uncertainty range that is strongly linked to the approximate description of hydraulic parameters \citep{Pappenberger_2008}, meteorological data \citep{Aronica_2012} and geographical data \citep{Morgan_2016}. In spite of the significant improvement in computational resources and the accuracy of numerical models, the turbulent nature of fluid mechanics equations limits the performance of hydraulic models. The need for realistic hydrodynamic flow simulation is beyond the abilities of deterministic forecast. As a consequence, model uncertainties have to be quantified. Recent advances in simulation and optimization of environmental systems have relied on increasingly detailed models, running many scenarios, in order to quantify these uncertainties \citep{Herman_2015,Teng_2017}.

According to \citet{solomatine_2008}, significant advances in the means of observing continental waters, especially through orbital equipment (SWOT satellite mission \citep{Morrow_2019}; Sentinel satellite missions (Copernicus program) \citep{Malenovsky_2012}), allows use of various types of data, with their associated analysis, for water resources management \citep{Knox_2019}. Thus, environmental modeling is increasingly complemented by data-driven models \citep{Kim_2010}. Moreover, by taking into account advances in information processing and data management, it becomes possible to reduce uncertainties and better understand the modeling process. Data-driven modeling is a paradigm shift for addressing many problems in science and engineering \citep{Montans_2019}.

To summarize, with the increase in computer resources, environmental modeling is evolving with the need to solve increasingly complex real-world problems involving the environment and its relationship to human systems and activities. If it is to break down research silos, and bring scientists, stakeholders and decision-makers together to solve social \citep{Daloglu_2014}, economic \citep{HAROU_2009} and environmental problems, an environmental modeling system should support IT tasks, such as aggregation of model components into functional units, component interaction and communication, parallel computing and cross-language interoperability \citep{David_2013,Laniak_2013}. These technical requirements are closely related to interoperability aspects, namely the capacity of the software to run and share information with other codes. Environmental system development requires both scientific understanding of the phenomena involved and software development proficiency.

This paper presents a new Application Programming Interface (API) of the open source and interoperable \telemacsystem (\url{http://svn.opentelemac.org/svn/opentelemac/}) to solve environmental problems \citep{Lacombe_2013,Goeury_2017}. Since 2017, the \telemacsystem has included in its distribution a generic API that allows it to expand its range of applications \citep{Audouin_2017}. With this newly implemented feature, the \telemacsystem can be easily coupled or integrated into higher-level platforms to model, study or simulate complex problems \citep{Goeury_2015}. 

To our knowledge, APIs for an environmental modeling system (with hydraulic (\mascaret 1D, \telemacdd, \telemacddd), water quality (\tracer, \waqtel), sediment transport (\courlis, \gaia), and wave hydrodynamics (\tomawac and \artemis)) are not yet widespread in the environmental community and are an original innovation. The present work uses as an example France's Gironde Estuary to demonstrate the operational applicability and outcomes of APIs. The port of Bordeaux in the Gironde Estuary faces many development challenges \citep{Klein_2018}. To satisfy demand for larger ships, while ensuring navigational safety, the evolution of water depth over time in the estuary needs to be predicted with a high degree of accuracy. To address this issue, the numerical results must be consistent with past observational data. Among other things, this process relies on calibration to determine ``empirical adequacy'' \citep{Oreskes_1994}. In particular, the calibration aims at simulating a series of reference events by adjusting some uncertain physically-based parameters until the comparison is as accurate as possible \citep{Dung_2011}. The objective of this work is to implement an efficient calibration algorithm, capable of processing measurement optimally, in order to estimate the partially known or missing parameters \citep{Nagel_2020}. Often performed, calibrating a hydrodynamic model is nevertheless a difficult task owing to the complexity of the flows and their interaction with the shoreline, the bathymetry, islands, etc. It is essential to understand, in depth, the relationship between the calibration of modeling parameters and the simulated state variables which are compared to the observations. In this context, a multivariate sensitivity analysis \cite{Lamboni_2014,Gamboa_2013} succeeded in identifying the most influential input parameters on the model results. Sensitivity analysis is associated to define the parameters to be calibrated using a physically based data-driven technique \citep{Carrassi_2018}.

This paper is organized as follows. Section \ref{sec:Method} includes a description of the main concepts of the \telemacsystem API. Section \ref{sec:architecture} presents the architecture of implemented APIs. Section \ref{sec:Estu_Gir} describes the application case. Section \ref{sec:calibration} deals with the calibration problem. Section \ref{sec:discussion} is the discussion and Section \ref{sec:conclu} is the conclusion.

\section{Method}
\label{sec:Method}

The \telemacmascaret hydro-informatic system, created in 1987, is an open source (\url{www.opentelemac.org}) integrated suite of solvers for use in the field of free-surface flows \citep{Hervouet_2007}. As displayed on Figure \ref{fng:TelMa_system}, it can carry out simulations of flows (\mascaret, \telemacdd and \telemacddd, {\color{gray} gray circle}), sediment transports (\courlis and \gaia, {\color{Apricot} orange circle}), waves (\artemis and \tomawac, {\color{YellowGreen} green circle}) and water quality (\tracer and \waqtel, {\color{RoyalPurple} purple circle}). Historically, from the original software solving physics equations (Saint-Venant: \mascaret and \telemacdd, Navier-Stokes: \telemacddd, elliptic mild slope equation: \artemis, simplified equation for the spectro-angular density of wave action: \tomawac, Exner: \courlis and \gaia, water quality processes: \tracer and \waqtel), the system has gradually evolved to the notion of hydro-informatics with a set of solvers dealing with large heterogeneous data and solving complex dependent problems (overlapping physical components in Figure \ref{fng:TelMa_system}). The various simulation components use high-end algorithms based on the finite element or finite volume method. The ability to study global, regional, or local scale problems by using the same system, makes the \telemacmascaret a useful tool for assessing the environmental state in the sea, estuaries, coastline and rivers.

\begin{figure}[!h]
\begin{adjustbox}{width=0.5\linewidth,center}
    \smartdiagramset{ 
       distance text center bubble=0.6cm, 
       bubble center node size=3cm, 
       bubble node size=3cm, 
       distance center/other bubbles=1.5cm, 
       bubble center node font=\scriptsize, 
        bubble node font=\scriptsize, 
%        bubble center node color=blue,
        distance center/other bubbles=0.5cm,
        bubble fill opacity =0.25,
%        set color list = {red,green, yellow} 
        set color list = {col3out,col4out, col8out} 
}% 
    \tikzset{
      bubble node/.append style={
        text width=2.6cm,
        align=center,let hypenation},
        bubble center/.append style={
        text width=0.5cm,
        align=center,let hypenation}
    }
\smartdiagram[bubble diagram]{
\textcolor{black}{\textnormal{
\begin{tabular}{l}
    Hydraulic:\\
    \mascaret (1D)\\
    \telemacdd (2D)\\
    \telemacddd (3D)\\ 
\end{tabular}}},
\textcolor{black}{\textnormal{
\begin{tabular}{l}
    Water quality:\\
    \tracer (1D)\\
    \waqtel (2D \& 3D)\\
\end{tabular}}},
\textcolor{black}{\textnormal{
\begin{tabular}{l}
    Waves:\\
    \artemis (2D)\\
    \tomawac (2D \& 3D)\\
\end{tabular}}},
\textcolor{black}{\textnormal{
\begin{tabular}{l}
    Sediment:\\
    \courlis (1D)\\
    \gaia (2D \& 3D)\\
\end{tabular}}}}
\end{adjustbox}
\caption{\telemacmascaret hydro-informatic system}
\label{fng:TelMa_system}
\end{figure}

According to \citet{Laniak_2013}, Integrated Environmental Modeling (IEM) highlights the importance of software development and sharing in their role as pluggable components of a larger ecosystem. The work presented here is part of this trend. It shows the implementation procedure for the component-based solver interoperability of the \telemacsystem. The API architecture, described in this paper, is directly applicable for any software regardless of its language since it is an easy task to build a wrapper in other programming language as presented in the following for Fortran and Python. The main difficult part is to split the software main program. When a solver is an interoperable component, it can be part of an assembly of elements in permanent interaction that work in a coordinated way. To be considered as an elementary reusable component, a computation model must: 

\begin{itemize}
    \item possess APIs, which can be viewed as new functionalities or entry points in the computational model to provide new services on demand;
    \item have a resolution algorithm comprising three main functionalities: problem initialization, simulation and finalization of the computation;
%    \item be usable under a dynamic library form;
    \item allow instantiation of several clearly separated and identifiable problems;
    \item be usable in dynamic library form.
\end{itemize}

In the following, the four points listed above are addressed.

\subsection{Design}

\begin{figure}[h!]
    \centering
    \begin{forest}
      for tree={
      font=\sffamily\bfseries,
      line width=1pt,
      %draw=col3out!100,
      ellip,
      align=center,
      child anchor=north,
      parent anchor=south,
      drop shadow,
      l sep+=5.5pt,
      edge path={
        \noexpand\path[rounded corners=5pt,thick, ->, shorten >=1pt,
        \forestoption{edge}]
          (!u.parent anchor) -- +(0,-5pt) -|
          (.child anchor)\forestoption{edge label};
        },
      where level={3}{tier=tier3}{},
      where level={0}{l sep-=5pt}{},
      where level={1}{
        if n={1}{
          edge path={
            \noexpand\path[rounded corners=5pt,thick, ->, shorten >=1pt,
              \forestoption{edge}]
              (!u.south) -| (.child anchor)\forestoption{edge label};
            },
        }{
          edge path={
            \noexpand\path[rounded corners=5pt,thick, ->, shorten >=1pt,
              \forestoption{edge}]
              (!u.south) -| (.child anchor)\forestoption{edge label};
            },
        }
      }{},
  }
    [\textcolor{black}{\textnormal{User calls}},name=root,rect, inner color=white, outer color=white
    [\textcolor{black}{\textnormal{Facade}},name=facade,
    [\textcolor{black}{\textnormal{module 1}},name=low1, rect, l=10mm, anchor=north west]
    [\textcolor{black}{\textnormal{module 2}},name=low2, rect,l=10mm, anchor=south west]
    [\textcolor{black}{\textnormal{$\cdots$}},name=low3, rect,l=12mm, anchor=south]
    [\textcolor{black}{\textnormal{module n}},name=low4, rect,l=12mm, anchor=base east]
    ]]
    \draw [line width=0.1mm,col3out ](-4,-4.1) -- (4,-4.1) -- (4,-1.2) -- (0.98,-1.2);
    \draw [line width=0.1mm,col3out ](-4,-4.1) -- (-4,-1.2) -- (-0.98,-1.2);
    \draw [col3out ] (0.7,-3.65) node [below] {\textnormal{Fortran sources of component-based solver}};
    \end{forest}
    \caption{Design Pattern “Facade”}
    \label{fig:facade}
\end{figure}
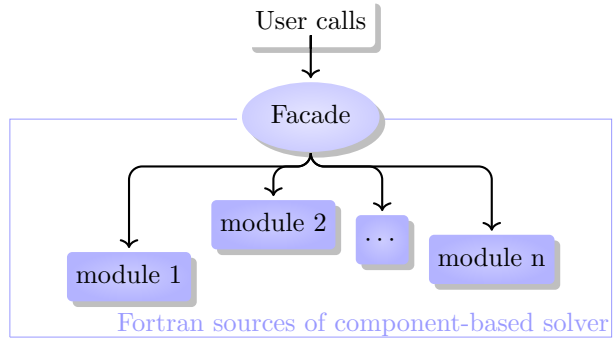

A computer programming model like the object-oriented programming (OOP) aims to organize the software design around data grouped into objects with attributes and methods. This help the developper to focus on the data rather than the logic required to manipulate them.
Most of the OOP concepts (encapsulation, interfaces, polymorphism, modularity, instantiation, \ldots) are derived from software engineering patterns but relate to the real world. As shown in Figure \ref{fig:facade}, the first concept used here is Encapsulation with a classic Design Pattern “Facade” \citep{Gamma_1994}. Since the \telemacsystem is written in Fortran $77$ and $90$, the design pattern has been adapted to this non object-oriented language as OOP is well-suited for the design of large and complex programs.

The main benefits of this approach are that it is easier to use and understand, limits the library user dependencies, includes integrity constraints and validation rules and avoids some unnecessary calculations \citep{Lacombe_2013}.

\begin{figure}[!h]
\centering
\includegraphics[width=0.75\linewidth]{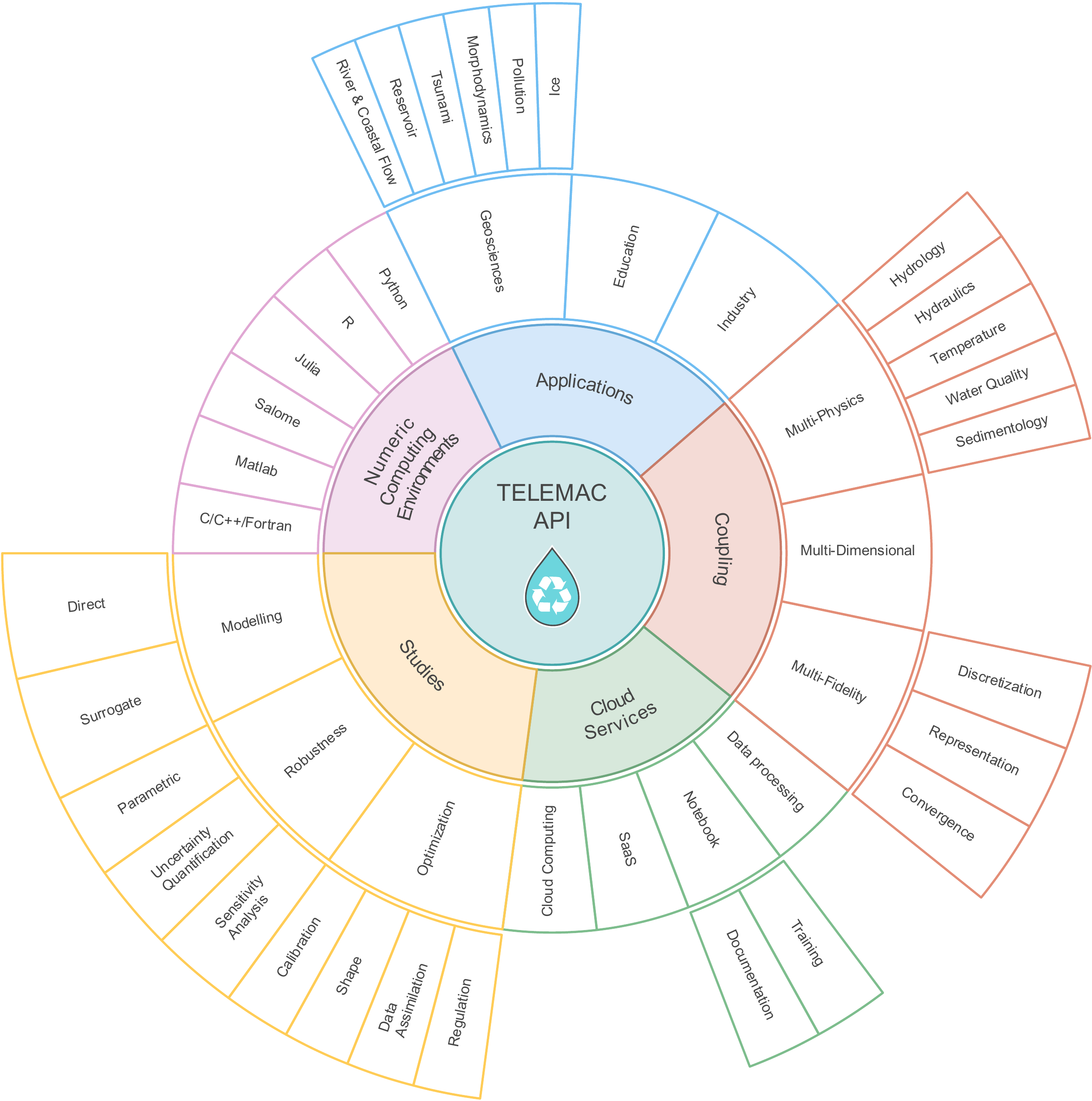}
\caption{Ecosystem of the \telemacmascaret API}
\label{fig:mind_map}
\end{figure}

Internal data has been accessed in order to expand the capabilities of the API. It is possible to examine all the information embedded in the calculation code and modify its structure (specifically the values, some metadata and properties) at runtime.
Pointers were used to access the simulation data. As Fortran lacks the pointer definition of other high-level languages such as C or C++, a combination of the \lstinline{POINTER} and \lstinline{TARGET} Fortran keywords with adequate updates were employed. For instance, a manually update must be done if a pointer on an \lstinline{ALLOCATABLE} variable is set before the dynamic allocation since the pointer is no longer valid. This ensures the correct state of the controlled variable. A component-based approach was adopted by having the same structure as all of the \telemacmascaret physical-dedicated solver, to allow them to be viewed as linkable. The execution of a simulation was partitioned to allow modification of a given parameter at runtime (for example, each time step). The flexibility that comes with interoperability allows deployment of the \telemacsystem in specific industrial applications and distribution of \telemacmascaret solvers as a part of complex modeling system (Applications, {\color[RGB]{101,181,240} blue part} in Figure \ref{fig:mind_map}). As mentioned by \citet{Gil_2019}, geosciences \textit{``will lead to a new generation of knowledge-rich intelligent systems that contain rich-know\-le\-dge and context in addition to data, enabling fundamentally new forms of reasoning, autonomy, learning, and interaction''}. As presented in Figure \ref{fig:mind_map}, the \telemacmascaret physical-dedicated components can be easily connected together to build multi-physics or multi-dimensional interactions \citep{Harpham_2014} (Coupling, {\color[RGB]{223,130,107} orange part}), inserted in optimization or uncertainty quantification processes (Studies, {\color[RGB]{255,197,76} yellow part}, see Section \ref{sec:calibration}), wrapped in advanced technology such as cloud-based solutions or embedded systems with data management (OpenMI \citep{Gregersen_2007} or Pynsim \citep{Knox_2018}, Cloud services, {\color[RGB]{114,181,129} green part}) and integrated in platforms (Salome \citep{Ribes_2007} or Palm \citep{Buis_2006} for example, Numeric computing environment, {\color[RGB]{220,157,205} pink part}).

\pagebreak

\subsection{Technology}
\label{subsec:technology}

To maintain efficient execution, a native code compiler (for example, GNU Compiler Collection, licensed under GNU GPL) running on most systems and hardware architectures, and supporting Fortran, is used and re\-co\-mmen\-ded. Moreover, the calculation code of the \telemacsystem numerically solves a model based on a system of partial differential equations (PDE). The computational solution relies on an implicit time scheme, removing stability constraints on the time frame, but leading to a linear-system solution of considerable size. The resulting increased complexity of the model structures poses strong limitations in terms of practical implementation and computational requirements. Consequently, High-Performance Computing (HPC) is necessary to tackle realistic applications. The solution deployed in the hydro-informatic system is to decompose the domain of computation. This idea of domain decomposition is to assign to each processor one part of the global domain over which it solves the fluid mechanics problem. The results of the other processors help in defining artificial boundary conditions arising from the partition. The assignments given to each processor are the same, but the data differ. Therefore, the \telemacsystem requires the use of an MPI library for communication between processors \citep{Clarke_1994}. 

It is possible to use MPI technology for the instantiation of several clearly separated and identifiable problems. This is because MPI is adapted for distributed memory systems. This method has been followed for the learning step of sensitivity analysis presented in Subsection \ref{subsec:sensitivity}.

\section{Architecture of API}
\label{sec:architecture}

The architecture is generic for all \telemacsystem components (Figure \ref{fng:TelMa_system}). For the sake of clarity, only an API example with the \telemacdd component is shown in the following sections.

\subsection{Fortran structure}

API’s main goal is to control the simulation while running a case. For example, it must allow the user to suspend the simulation at any time step, retrieve some variable values, and possibly change them before resuming the calculation. To make this possible, a Fortran structure called ``instance'' is used in the API. The instance structure gives direct access to the physical memory of variables, and therefore allows control of the variable. Furthermore, based on decomposition of the main \telemacsystem subroutines in three main functionalities (initialization, simulation of one time step and finalization), it is possible to run a hydraulic case for an indefinite sequence of time steps.

\begin{figure}[!ht]
\centering
\begin{adjustbox}{}
\begin{tikzpicture} [
    auto,
    decision/.style = { diamond, draw=col3out!100, thick, fill=col3out!20,
                        text width=5em, text badly centered,
                        inner sep=1pt, rounded corners },
    block/.style    = { rectangle, draw=col3out!100, thick, 
                        fill=col3out!20, text width=15em, text centered,
                        rounded corners, minimum height=2em },
    block2/.style    = { rectangle, draw=white, thick, 
                         fill=white, text width=15em, 
                         rounded corners, minimum height=2.45em },
    line/.style     = { draw,thick, shorten >=1pt, -> },
    line2/.style     = { draw,thick, shorten >=1pt, snake=triangles, segment length=4pt,transform canvas={xshift=4pt} },
  ]
  % Define nodes in a matrix
  \matrix [column sep=5mm, row sep=3mm] {
                    & \node [text centered] (in) {\textcolor{black}{\textnormal{Input parameter vector ${\bf{X}}=\left(X_1,...,X_p\right)'$}}} ;            & \\
                    & \node [coordinate](null0) {};                   & \\
                    & \node [block] (inst) {\textcolor{black}{\textnormal{Configuration setup (initialize instance ID and listing output)}}};       & \node [block2] (inst2) {\textcolor{black}{\lstinline{CALL RUN_SET_CONFIG_T2D}}}; \\
                    & \node [block] (setcase) {\textcolor{black}{\textnormal{Set study physical and numerical parameter}}};                   & \node [block2] (setcase2) {\textcolor{black}{\lstinline{CALL RUN_READ_CASE_T2D}}}; \\
                    & \node [block] (initstate) {\textcolor{black}{\textnormal{Set the initial condition of computation ($t_0$)}}};         & \node [block2] (initstate2) {\textcolor{black}{\lstinline{CALL RUN_ALLOCATION}\\\lstinline{
                    CALL RUN_INIT}}};\\
                    & \node [block] (set) {\textcolor{black}{\textnormal{Change the parameter values according to ${\bf{X}}$}}};   & \node [block2] (set2) {\textcolor{black}{\lstinline{CALL SET_DOUBLE}}};\\
                    & \node (null1) {};                                    & \\
                    & \node [block] (run) {\textcolor{black}{\textnormal{Run one time step of the solver ($t_i$)}}};   & \node [block2] (run2) {\textcolor{black}{\lstinline{CALL RUN_TIMESTEP_T2D}}};\\
                    & \node [block] (get) {\textcolor{black}{\textnormal{Get the output desired values $O(t_i)$}}};   & \node [block2] (get2) {\textcolor{black}{\lstinline{CALL GET_DOUBLE}}};\\
    \node(null3){}; & \node [decision] (timeloop) {\textcolor{black}{\textnormal{$i <= n$}}};                &  \\
                    & \node [block] (finalize) {\textcolor{black}{\textnormal{Conclude solver computation (deallocation,  and so on)}}}; & \node [block2] (finalize2) {\textcolor{black}{\lstinline{CALL RUN_FINALIZE_T2D}}}; \\
                    & \node [coordinate] (null2) {};                   & \\
                    & \node [text centered] (out) {\textcolor{black}{\textnormal{$O(t)=M\left({\bf{X}},t\right)$}}};          & \\
  };
  % connect all nodes defined above
  \begin{scope} 
    \path [line] (in)       --    (inst);
    \path [line] (inst)     --    (setcase);
    \path [line] (setcase) --    (initstate);
    \path [line] (initstate) --    (set);
    \path [line] (set) --    (run);
    \path [line] (run) --    (get);
    \path [line] (get) --    (timeloop);
    \path [line] (timeloop) --    node [near start] {no} (finalize);
    \path [line] (timeloop) --++   (-3,0)  node [near start] {yes} |- (null1);
    \path [line] (finalize) --    (out);
    \path [line2] (inst)    --    (inst2);
    \path [line2] (setcase)     --    (setcase2);
    \path [line2] (initstate) --    (initstate2);
    \path [line2] (set) --    (set2);
    \path [line2] (run) -- (run2);
    \path [line2] (get) -- (get2);
    \path [line2] (finalize) -- (finalize2);
    
  \end{scope}
\end{tikzpicture}

\end{adjustbox}

\caption{Diagram of study workflow (Function $M\left({\bf{X}},t\right)$ in Figure \ref{fng:model_cali}) and corresponding Fortran API calls}
\label{fig:telapy_diag}
\end{figure}
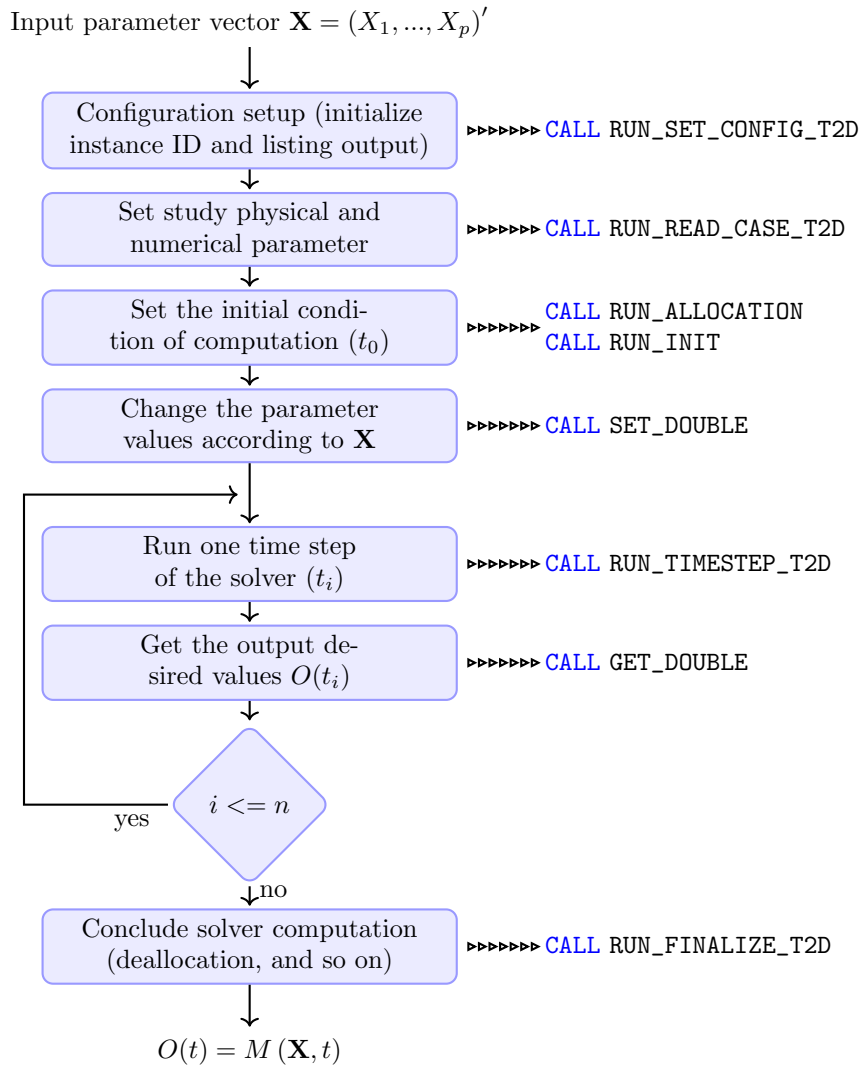

Figure \ref{fig:telapy_diag} describes an example of API workflow where ${\bf{X}}$ are input parameters and $O(t)$ is information extracted at given times. For instance, in the framework of the Gironde Estuary application case, ${\bf{X}}$ represents parameters to be estimated / calibrated (friction coefficients and tidal boundary condition parameters) and $O(t)$ is the free surface elevation every minute at the measurement stations (see Section \ref{sec:Estu_Gir}).

For each action defined above, the identity number of the instance is used as an input argument allowing all computation variables to be linked with the corresponding instance pointers. These actions must be done in chronological sequence in order to ensure proper execution of the computation in the API main program. In Fortran and for the shallow water code \telemacdd, it will begin with the call of an initialization subroutine:

\begin{lstlisting}[language=Fortran]
CALL RUN_SET_CONFIG_T2D(ID, LU, LNG, COMM, IERR)
\end{lstlisting}
where all parameters are scalars of type integer and \lstinline{ID} is an output value giving the instantiation number. The input parameter \lstinline{LU} is used to redirect the standard output, \lstinline{LNG} has only two possible values for the choice of the english or french language, and \lstinline{COMM} is the MPI communicator for the distributed-memory parallelism. The output \lstinline{IERR} has to be considered with care as a non-null value stands for the occurrence of errors. A systematic test on this return value is recommended, for instance one may want to stop the program accordingly:
\begin{lstlisting}[language=Fortran]
IF (IERR.NE.0) EXIT IERR
\end{lstlisting}
After the initialization phase, the model has to be read from files with the following instruction:
\begin{lstlisting}[language=Fortran]
CALL RUN_READ_CASE_T2D(ID, CAS_FILE, DICO_FILE, INIT, IERR)
\end{lstlisting}
The name of the steering file is given with \lstinline{CAS_FILE} as a character string, it stores several input data of the model with key-value pairs. Each possible key is set with a dictionary file whose name is a character string hold by \lstinline{DICO_FILE}. A true value for the logical input \lstinline{INIT} will initialize a common computational kernel of all the modules. After this step, a dynamic allocation of arrays in memory is necessary followed by a state initialization of the physics:
\begin{lstlisting}[language=Fortran]
CALL RUN_ALLOCATION_T2D(ID, IERR)
CALL RUN_INIT_T2D(ID, IERR)
\end{lstlisting}
All accessible parameters are listed in predefined keyword lists (see Appendix \Ref{appendix_apivar} for \telemacdd). For example, the number of time steps in the simulation is named \lstinline{MODEL.NTIMESTEPS} and its value is obtained and stored in the variable \lstinline{NBSTEPS} with the following call:
\begin{lstlisting}[language=Fortran]
CALL GET_INTEGER(ID, "T2D", "MODEL.NTIMESTEPS", NBSTEPS, 0, 0, 0, IERR)
\end{lstlisting}
Input parameter can be defined here using the following call:
\begin{lstlisting}[language=Fortran]
CALL SET_DOUBLE(ID, "T2D", "MODEL.SEALEVEL", SEALEVEL, 0, 0, 0, IERR)
\end{lstlisting}
where the sea level value in the computation (accessed through the keyword \lstinline{MODEL.SEALEVEL}) is set to the value of the variable \lstinline{SEALEVEL}.

Now the simulation can be done with a loop on the number of time steps while storing all the results of an evolution of the water depth:
\begin{lstlisting}[language=Fortran]
DO I = 1, NBSTEPS
  CALL RUN_TIMESTEP_T2D(ID, IERR)
  CALL GET_DOUBLE(ID, "T2D", "MODEL.WATERDEPTH", WATERDEPTH(I), 42, 0, 0, IERR) 
ENDDO
\end{lstlisting}
\lstinline{WATERDEPTH} is a one-dimensional array of size \lstinline{NBSTEPS}. It is used here to store the water depth values at node number $42$ of the \telemacdd triangular mesh. Finally, the Fortran program can end with the instance deletion thus freeing the memory used:
\begin{lstlisting}[language=Fortran]
CALL RUN_FINALIZE_T2D(ID, IERR)
\end{lstlisting}

Additional functions are available to handle parallelism (see Subsection \ref{subsec:technology} for more details on parallelism in the \telemacsystem). The data structure in the current version (official version 8.2) does not yet allow for multiple instantiations, as all the instances point to the same memory area. This constitutes a future development to improve API capacity. However, instantiation associated with the processor communication function overcomes this limitation by ensuring the possibility of several clearly separated and identifiable problems.

\subsection{Python wrapping}

The use of API is not limited to Fortran programing but can also be used by a scripting language. To this end, a Python wrapper is also available. It is relatively easy to use the Fortran API routines directly in Python using the f2py tool of the Python Scipy library \citep{Peterson_2009}. This tool will make it possible to compile Fortran code for use in Python. The only limitations are on the type of arguments of the functions to wrap. Python is a portable, dynamic, extensible, free language, which allows (without requiring) a modular approach and object-oriented programming \citep{Rossum_2009}. In addition to the benefits of this programming language, Python offers a large number of interoperable libraries for scientific computing, image processing, data processing, machine learning and deep learning. The link between various interoperable libraries with \telemacsystem APIs allows the creation of an ever-more efficient calculation chain capable of responding finely to various complex problems such as the application case presented in this article (see Section \ref{sec:Estu_Gir}). Moreover, the Python scripting language makes it possible to implement a wrapper in order to provide user-friendly functioning of the Fortran API. Thus, a Python overlay was developed to encapsulate and simplify the different API calls. 

\begin{figure}[h!]
\begin{adjustbox}{width=0.65\linewidth,center}
%\hspace{-7.cm}
 \begin{forest}
  for tree={
      font=\rmfamily\bfseries,
      line width=1pt,
      ellip,
      align=center,
      child anchor=north,
      parent anchor=south,
      drop shadow,
      l sep+=12.5pt,
      edge path={
        \noexpand\path[rounded corners=5pt,thick, ->, shorten >=1pt,
          \forestoption{edge}]
          (!u.parent anchor) -- +(0,-5pt) -|
          (.child anchor)\forestoption{edge label};
        },
      where level={3}{tier=tier3}{},
      where level={0}{l sep-=15pt}{},
      where level={1}{
        if n={1}{
          edge path={
            \noexpand\path[rounded corners=5pt,thick, ->, shorten >=1pt,
              \forestoption{edge}]
              (!u.west) -| (.child anchor)\forestoption{edge label};
            },
        }{
          edge path={
            \noexpand\path[rounded corners=5pt,thick, ->, shorten >=1pt,
              \forestoption{edge}]
              (!u.east) -| (.child anchor)\forestoption{edge label};
            },
        }
      }{},
  }
  [\textcolor{black}{TelApy module}, inner color=col1in, outer color=col1out
    [\textcolor{black}{Fortran API}, inner color=col3in, outer color=col3out
      [\textcolor{black}{Main function of API}\\\textcolor{black}{module api\_interface}
        [\textcolor{black}{TELEMAC-2D solver}, rect, name=sse1
        [\textcolor{black}{instantiation module}\\\textcolor{black}{api\_instance\_t2d}, rect, name=sse2
        [\textcolor{black}{Variable control}\\\textcolor{black}{api\_handle\_var\_t2d}, rect, name=sse3
        [\textcolor{black}{Computation control}\\\textcolor{black}{api\_run\_t2d}, rect, name=sse4]
        ]
        ]
        ]
        [\textcolor{black}{TELEMAC-3D solver}, rect, name=sse5
        [\textcolor{black}{instantiation module}\\\textcolor{black}{api\_instance\_t3d}, rect, name=sse6
        [\textcolor{black}{Variable control}\\\textcolor{black}{api\_handle\_var\_t3d}, rect, name=sse7
        [\textcolor{black}{Computation control}\\\textcolor{black}{api\_run\_t3d}, rect, name=sse8]
        ]
        ]
        ]
        [\textcolor{black}{solver ...}, rect, name=sse9
        ]
      ]
    ]
    [\textcolor{black}{Python wrapper}, inner color=col2in, outer color=col2out
      [\textcolor{black}{Class api\_module}, inner color=col4in, outer color=col4out
      [\textcolor{black}{Class t2d}, inner color=col4in, outer color=col4out
        [\textcolor{black}{Specific function for }\\\textcolor{black}{TELEMAC-2D}, rectpy, name=pyt2D
        ]
      ]
      [\textcolor{black}{Class t3d}, inner color=col4in, outer color=col4out
        [\textcolor{black}{Specific function for }\\\textcolor{black}{TELEMAC-3D}, rectpy, name=pyt3D
        ]
      ]
      [\textcolor{black}{Class ...}, inner color=col4in, outer color=col4out
      [\textcolor{black}{Specific function for }\\\textcolor{black}{...}, rectpy, name=pyunk2D]
      ]
      ]
    ]
  ]
  \begin{scope}[color = col3out!100, rounded corners = 5pt,
    >={Stealth[length=10pt]}, line width=1pt, ->];
%    \draw (sse4.south) -- (us.north -| sse4.south);
%    \draw (sse8.south) -- (us.north -| sse8.south);
%        \draw (sse9.south) -- (us.north -| sse9.south);
%   \coordinate (c1) at ($(sse4.south)!2/5!(sse8.south)$);
%    \coordinate (c2) at ($(sse8.south)!2/5!(sse9.south)$);
%    \draw (sse8.south) -- +(0,-10pt) -| (us.north -| c1);
%    \draw (sse9.south) -- +(0,-10pt) -| (us.north -| c2);
  \end{scope}
%  \draw [line width=0.5mm,col3out ](-4,-4.1) -- (4,-4.1) -- (4,-1.2) -- (0.98,-1.2);
%%%Arrow between python and fortran part  \draw [<->, > = stealth,line width=1.5mm, col1out] (-3.3, -1.2) -- (2.8, -1.2) node [] {} ;
\end{forest}
\end{adjustbox}
\caption{Overview of TelApy module}
\label{fng:api_over}
\end{figure}
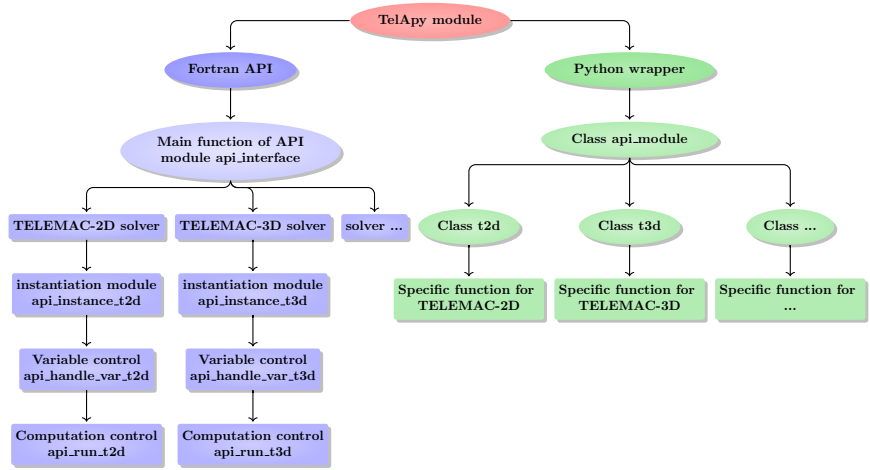

For instance, a Python ``get'' function can encapsulate all API Fortran type-dedicated function get\_\textit{type} (where \textit{type} can be replaced by Double, Integer, Boolean and String). File management is also made easier with Python (copy, mo\-ve, \ldots, and reducing the number of arguments). This Python wrapping of \telemacsystem API constitutes a package called TelApy distributed in the official version of the hydro-informatic system (Figure \ref{fng:api_over}).

The TelApy package is provided with a tutorial for people who want to run the \telemacsystem physical components in an interactive mode with the help of the Python scripting language. The script corresponding to the application case of Section \ref{sec:Estu_Gir} is presented in Appendix \Ref{appendix_telapy}.

\subsection{Summary}

A single coherent Fortran structure has been deployed in \telemacmascaret in order to ensure interoperability of the system for each of its solvers. This is based on three main functionalities: instantiation, variable values and computation control. Firstly, instantiation associated with the processor communication function ensures the possibility of a simultaneous existence of several clearly separated and identifiable problems. Secondly, based on variable and computation control, memory access is permitted during simulation, allowing new services to be provided on demand. Finally, the technology allows dynamic compilation of each physical-based component model. All of these features enable the \telemacsystem to operate with other existing or future codes without restricting access or implementation. It therefore becomes natural to drive these APIs using Python scripting language. As mentioned by \citet{Knox_2018}, this scripting language offers numerical and scientific libraries and is already a recognized tool in environmental resource modeling.

Finally, the Python wrapping allows use of the \telemacsystem for a wide range of studies, such as optimization, coupling and uncertainty quantification.

The Application Programing Interface (API) framework described in this work (Fortran APIs and its Python wrapper) is available for download and distributed in \telemacsystem (\url{www.opentelemac.org}).

\section{Gironde Estuary application case}
\label{sec:Estu_Gir}

\begin{figure}[!h]
\centering
\includegraphics[width=0.65\linewidth]{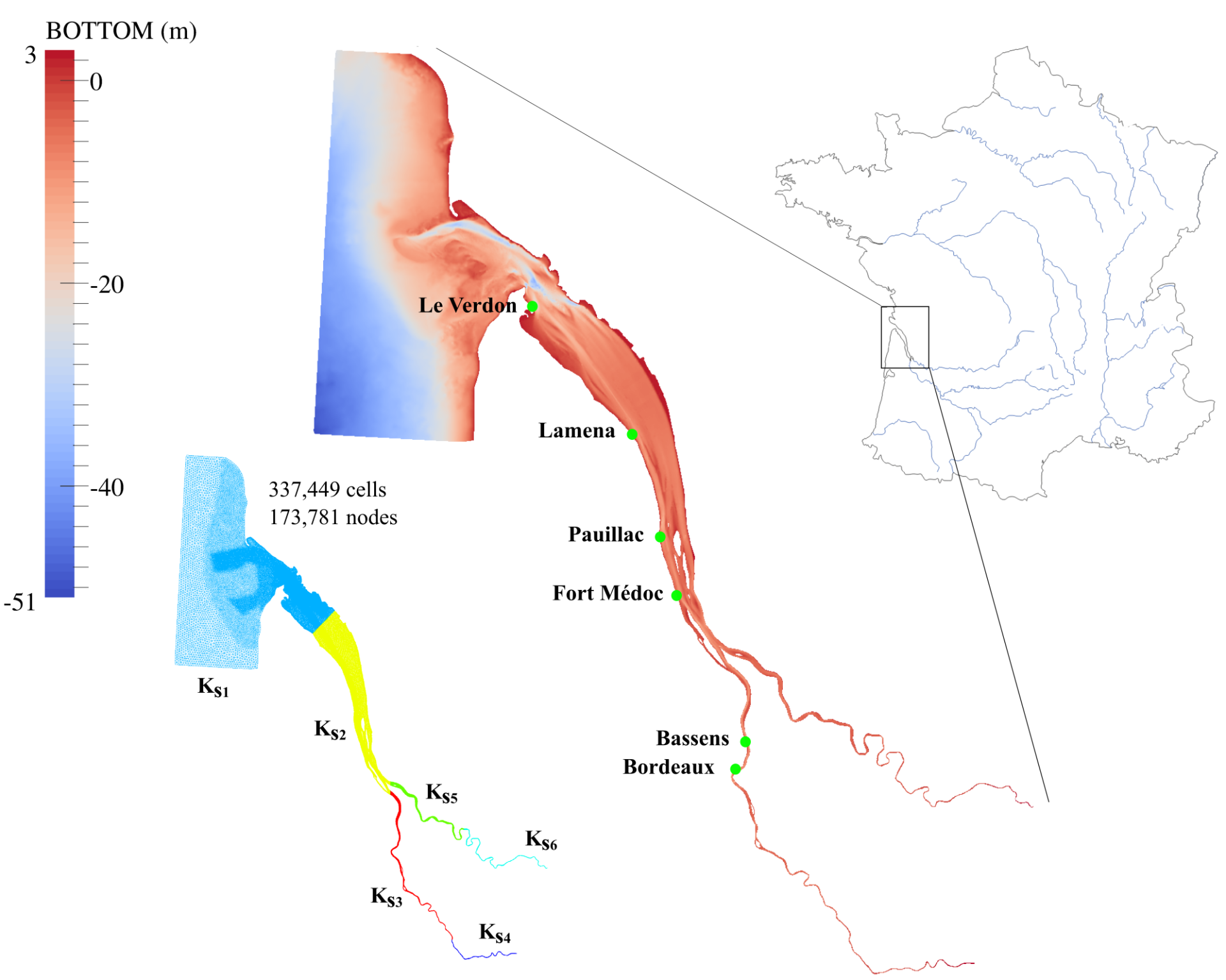}
\caption{Gironde Estuary. Location of observation stations (\textcolor{green}{$\bullet$}), bathymetry, friction zones ($K_s$ denotes Strickler coefficient) and mesh size \citep{Zaoui_2019}}
\label{fig:estu_gir}
\end{figure}

The hydrodynamic model \telemacdd was used to solve the shallow water equations (see Section \ref{sub:hydro}) in the real case of the Gironde Estuary in southwestern France (Figure \ref{fig:estu_gir}). This important large-scale estuary involves many economic and environmental considerations. Several hydrodynamic and morphodynamic \telemacdd studies of the estuary have been carried out for the last decade \cite{Huybrecht_2012,Villaret_2010}. In these studies, the importance of the calibration phase was systematically emphasized to achieve operational per\-for\-man\-ce.

Model calibration aims at simulating a series of reference events by adjusting some uncertain physically based parameters in order to minimize deviation between measured and computed values of variables of interest. This process of parameter estimation, a subset of the so-called inverse problems, consists in evaluating the underlying input data of a problem from its solution. For free-surface flow hydraulics, parameters that are often unknown or difficult to assess include initial state, bathymetry, bed friction and model boundary conditions. In this work, the bathymetry and initial state are considered as known. In fact, a spin-up of 12 hours was used to generate realistic initial condition. Bathymetry uncertainty was not considered in this study and could constitute an approximation in particular on current speed \cite{Cea_2012}. A primary challenge of integrating bathymetry in the calibration process is the spatial structure characterization of its uncertainty in relation to channel morphology \cite{Legleiter2011}. This was not studied here and will be the topic of a future study.
Consequently, both bed roughness and model boundary conditions are considered in the calibration process. As mentioned by \citet{Williams2017}, the hydrodynamic bed friction is a primary calibration variable for all coastal and estuarine models. It is also essential for modeling other processes accurately such as sediment transport and wave attenuation. Concerning the model boundary conditions, only offshore boundary condition is considered in the calibration process as it is derived from a coarser-scale model and upstream boundary conditions is coming from measurement. The transfer of information between a large scale model and the boundaries of a more local model generally requires empirical adequacy: sea level could need to be corrected to account for seasonal variability (effect of thermal expansion, salinity variations, air pressure, etc.) in addition to long-term sea level rise resulting from climate change and differences in tidal amplitude could be attributable to meteorological effects (storm and surge (atmospheric and wave setup)) \cite{Idier_2019}. Although this paper focuses on friction and tidal parameters, all \telemacdd variables can be changed by the API, as presented in Table \ref{Tab:api_var}.

\subsection{Numerical configuration and available data}

The Gironde is a navigable estuary in southwest France formed by the confluence of the Garonne and Dordogne ri\-vers just downstream of the center of Bordeaux. The Gironde Estuary is the largest estuary in western Europe. The hydraulic model covers approximately $195$ $\text{km}$ between the fluvial up\-stre\-am and the maritime downstream boundary conditions, for an area of around $635$ $\text{km}^2$. The finite element mesh is composed of $173,781$ nodes (Figure \ref{fig:estu_gir}). The mesh size varies from $40$ $\text{m}$ within the area of interest, the navigation channel, to $750$ $\text{m}$ offshore (western and northern sectors of the model). As shown in Figure \ref{fig:estu_gir}, six friction areas are considered in the hydraulic model.

The boundary condition along the marine border of the model was set using depth-averaged velocities and water levels from the Legos numerical model TUGO dataset (46 harmonic constants). Surge data, describing the difference between the tidal signal and the observed water level, are taken into account using a data file generated from the Hycom2D model of the SHOM \citep{Chassignet_2007}. Surface wind data are also considered to simulate the flow under windy conditions. Ti\-m\-e-evolution flow discharge hydrographs are imposed up\-str\-eam of the Gironde Estuary model on the Garonne and Dordogne rivers. The ability of the numerical model to reproduce complex physical processes such as sediment transport is presented in \cite{ORSEAU_2020}.

The free surface flow was continuously measured (every minute) at the Verdon, Richard, Lamena, Pauillac, Fort Médoc, Ambes, The Marquis, Bassens and Bordeaux locations (Figure \ref{fig:estu_gir}). For this study, observation results are used over a $36$ hour period from August $12$ to $14$, $2015$.

\subsection{Hydrodynamic model}
\label{sub:hydro}

The \telemacdd code solves the 2D depth-averaged free surface flow equations, also known as shallow water equations (Eqs. \ref{eq:mass}-\ref{eq:mvty}).

\begin{align}
\label{eq:mass}
\frac{\partial h}{\partial t} + \frac{\partial}{\partial x}\left(h u\right) + \frac{\partial}{\partial y}\left(h v\right)=& 0\\
\label{eq:mvtx}
\frac{\partial h u}{\partial t}+ \frac{\partial}{\partial x}\left(h u u\right)+\frac{\partial}{\partial y}\left(h u v\right)=&-g h\frac{\partial Z_s}{\partial x}+h F_x+\nabla\cdot\left(h\nu_e{\bf{\nabla}}\left(u\right)\right)\\
\label{eq:mvty}
\frac{\partial h v}{\partial t}+ \frac{\partial}{\partial x}\left(h u v\right)+\frac{\partial}{\partial y}\left(h v v\right)=&-g h\frac{\partial Z_s}{\partial y}+h F_y+\nabla\cdot\left(h\nu_e{\bf{\nabla}}\left(v\right)\right)
\end{align}

where $x$ and $y$ are the horizontal Cartesian coordinates, $t$ the time, $u$ and $v$ the components of the depth-averaged velocity, $h$ the water depth, $\nu_e$ an effective diffusion representing depth-averaged turbulent viscosity $\nu_t$ and dispersion, $Z_s$ the free surface elevation, $g$ the gravitational acceleration, $F_x$ and $F_y$ refer to the friction force (see Section \ref{subsec:friction}).

\telemacdd solves the previous equation system using the finite element method on a triangular element mesh. The main results at each node of the computational mesh are the water depth and the depth-averaged velocity components. \telemacdd can take into account propagation of long waves, including non-linear effects, bed friction, effect of the Coriolis force, effects of meteorological phenomena (e.g., atmospheric pressure, rain or evaporation, and wind), turbulence, supercritical and subcritical flows, influence of horizontal temperature and salinity gradients on density and dry area in the computational domain, amongst other processes \citep{Hervouet_2007}.

\subsubsection{Friction coefficient}
\label{subsec:friction}

The friction term in the momentum part of the shallow water equations is treated in a semi-implicit form within \telemacdd. The two components of friction force are given in Eq. \ref{eq:fric}.

\begin{equation}
    \left\{
    \begin{array}{ll}
        F_x =& -\frac{u}{2h}C_f\sqrt{u^2+v^2} \\
        F_y =& -\frac{v}{2h}C_f\sqrt{u^2+v^2}
    \end{array}
\right.
\label{eq:fric}
\end{equation}

where  $C_f$ is a dimensionless friction coefficient.

The roughness coefficient often takes into account friction caused by walls on the fluid or other phenomena such as turbulence. Empirical or semi-empirical formulas are used for calculating $C_f$ \citep{Morvan_2008}. In the present work, $C_f$ is given by the widely used Strickler formulas (Eq. \ref{Eq:Strickler_formula}). The Strickler coefficient ($\text{m}^{1/3}\text{s}^{-1}$) is a calibration parameter of the modeling system, to be adjusted according to field data (e.g., usual measured water levels).

\begin{equation}
C_{f}=\frac{2g}{h^{1/3}K_s^{2}}
\label{Eq:Strickler_formula}
\end{equation}

Generally, the friction coefficient is contained in an interval bounded by physical values, as it results from different contributions (e.g., skin friction, bed form dissipation, etc.). In this study, the Strickler bounds are taken as large as possible. In fact, the lower bound of the Strickler coefficient is set to $5$ according to U.S.F.H.A \cite{united1984guide} (rough bed surface) and the upper one is set to $115$ (extreme smooth bed surface).

In the present API context, the friction coefficient is called \lstinline{MODEL.CHESTR} and in the study corresponds to the Strickler coefficient $K_{si}$ (where $i$ denotes the considered friction area).

\subsubsection{Tidal amplification parameter}
\label{subsec:tidal}

Tidal characteristics are imposed using a database of harmonic constituents to force the open boundary conditions. For each harmonic constituent, the water depth $h$ and components of velocity $u$ and $v$ are calculated, at point $P$ and time $t$ by Eq. \ref{eq:harmonic}.

\begin{equation}
    \left\{
    \begin{array}{ll}
        f\left(P,t\right) =& \sum_{i} f_i\left(P,t\right)\\
        f_i\left(P,t\right)=&f_i\left(t\right)A_{f_i}\left(P\right)cos\left[\frac{2\pi t}{T_i}-\varphi_{f_i}\left(P\right)+\varphi_i^0+l_i\left(t\right)\right]
    \end{array}
\right.
\label{eq:harmonic}
\end{equation}

where $f$ is either the water depth $h$ or one of the components of velocity $u$ or $v$, $i$ refers to the considered constituent, $T_i$ is the period of the constituent, $A_{f_i}$ is the amplitude of water depth or one of the horizontal components of velocity, $\varphi_{f_i}$ is the phase, $f_i \left(t\right)$ and $l_i \left(t\right)$ are the nodal factors and $\varphi_i^0$ is the phase at the original time of the simulation. 
The water level and velocities of each constituent are then submitted to obtain the water depths and velocities for the open boundary conditions (Eq. \ref{eq:tidalcl}).

\begin{equation}
    \left\{
    \begin{array}{ll}
        h = & \alpha\sum_{i}h_{i}-Z_b+Z_{ref}-\gamma\\
        u = & \beta\sum_{i}u_i\\
        v = & \beta\sum_{i}v_i
    \end{array}
\right.
\label{eq:tidalcl}
\end{equation}

where $Z_b$ is the bottom elevation and $Z_{ref}$ the mean reference level. In Eq. \ref{eq:tidalcl}, the tidal amplitude multiplier coefficients of tidal range and velocity, respectively $\alpha$ and $\beta$, at boundary locations and the sea level correction $\gamma$ are assumed to be the tidal calibration parameters \citep{Pham_2012}.

The sea level correction parameter is assumed, following expert opinion, to be contained in an interval of plus or minus a meter. The interval of the weighting coefficients of tidal range and velocity is set to $[0.8, 1.2]$ corresponding to a $20$\% margin of the initial value.

In the API framework, the sea level correction variable $\gamma$ is designated as \lstinline{MODEL.SEALEVEL}, and the weighting coefficients of tidal range $\alpha$ and velocity $\beta$ are denoted \lstinline{MODEL.TIDALRANGE} and \lstinline{MODEL.VELOCITYRANGE}, respectively.

\subsection{Summary of parameter range variation}
\label{sub:range_var}

All the model input parameters and their associated probability distribution are summarized in Table \ref{Tab:input_var}.

%%%\begin{longtable}{|p{.35\textwidth} | p{.1\textwidth}p{.15\textwidth}p{.2\textwidth}|}
\begin{longtable}{p{.35\textwidth} | ccc}
\hline
\textbf{Variable name} & \textbf{Nature} & \textbf{Variation interval} & \textbf{Probability distribution} \tabularnewline
\hline
\hline
Fiction coefficient per area %($\text{m}^{1/3}\text{s}^{-1}$) & Real scalar &
($m^{1/3}s^{-1}$) & Real scalar &$\left[5.,115.\right]$ & Uniform\\

Sea level correction parameter ($m$)& Real scalar & $\left[-1.,1.\right]$ & Uniform\\

Tidal range coefficient ($-$)& Real scalar & $\left[0.8,1.2\right]$ & Uniform\\

Velocity range coefficient ($-$)& Real scalar & $\left[0.8,1.2\right]$ & Uniform\\
\hline
\hline
\caption{Input parameters and associated probabilistic model for the Gironde Estuary application}
\label{Tab:input_var}
\end{longtable}

\section{Calibration problem}
\label{sec:calibration}

The inverse problem of calibration can be understood as the computation of the \textit{posterior} distribution $\pi\left({\bf{X}}|{\bf{Y}}\right)$ (``Mo\-del Cal\-ibra\-tion Algorithm step'', Figure \ref{fng:model_cali}, as presented in appendix \ref{appendix_bayesian}), where model parameters constitute the p-components of the parameter control vector ${\bf{X}} = \left(X_1,...,X_p\right)'\in\mathbb{R}^p$ composed of independent variables defined on some probability space $\left(\Omega,A,\mathbb{P}\right)$ (with $\Omega$ as a sample space, $A$ the $\sigma$-algebra of events and $\mathbb{P}$ the probability measure), ${\bf{Y}}\in\mathbb{R}^m$ is the observation vector, also defined on a probability space, around the unknown parameter vector ${\bf{X}}$ and $\mathbb{R}^m$ is the observation space defined by Eq. \ref{eq:observation_space}.
 
 \begin{equation}
     {\bf{Y}}=G\left({\bf{X}}\right)+\epsilon_m
     \label{eq:observation_space}
 \end{equation}
 
where $G:\mathbb{R}^p\longrightarrow\mathbb{R}^m$ is a vector-valued function of vector ${\bf{X}}$ and $\epsilon_m\in\mathbb{R}^m$ is an observable measurement noise such as $\mathbb{E}\left(\epsilon_m\right)=0$ and $\mathcal{R} = cov(\epsilon_m)=\mathbb{E}\left(\epsilon_m\epsilon_m'\right)\in\mathbb{R}^{m\times m}$ and identified as a multivariate normal distribution, $\epsilon_m \sim \mathbb{N}\left(0,\mathcal{R}\right)$. As a reminder, in the Gironde Estuary application case, the parameter control vector ${\bf{X}}$ is composed of the friction and tidal parameters and the observation parameter ${\bf{Y}}$ is composed of water levels measured every minute at an observation station. Note that the operator $G$ enabling the passage from the parameter space (where vector ${\bf{X}}$ lives) to the observation space (where vector ${\bf{Y}}$ lives) represents a call to the API study workflow defined in Figure \ref{fig:telapy_diag}. As \telemacdd is a discrete time dynamic model, the output ${\bf{O}}$ is composed of scalar output (water elevation at observation stations interpolated from \telemacdd computational nodes given at $t\in \mathbb{T}=\left[1,..,T\right]$).

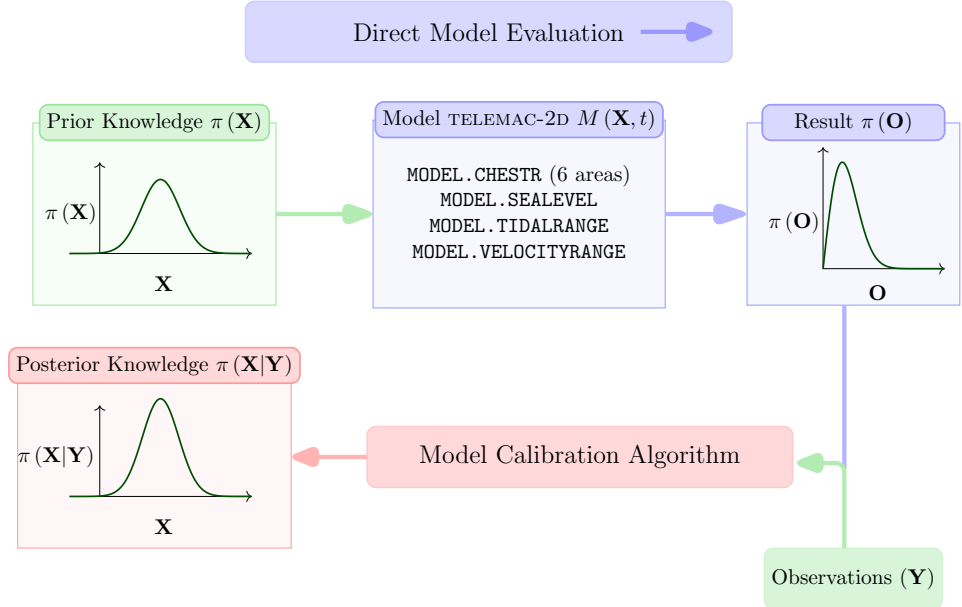
\begin{figure}[h!]
\begin{adjustbox}{width=0.8\linewidth,center}
% Define the layers to draw the diagram
\pgfdeclarelayer{background}
\pgfdeclarelayer{foreground}
\pgfsetlayers{background,main,foreground}
\begin{tikzpicture}[ampersand replacement=\&,font=\rmfamily]
%     \draw[help lines](0,-5) grid (10,5); 
\begin{scope}[xshift=-3cm, yshift=5cm] (model)
\node[rectangle,draw=col3out!100,fill=col3in!10,inner sep=10pt, inner ysep=20pt, minimum width=3cm,minimum height=2cm,text justified,text=black] (mabox)
{ \begin{tabular}{c} \text{\lstinline{MODEL.CHESTR}} (6 areas) \\  \text{\lstinline{MODEL.SEALEVEL}}\\\text{\lstinline{MODEL.TIDALRANGE}}\\\text{\lstinline{MODEL.VELOCITYRANGE}}\end{tabular} };
% le titre placé grace au label du noeud
\node[rectangle,draw=col3out!100,rounded corners,thick,fill=col3in!50,text=black]at (mabox.north)
{Model \telemacdd $M\left(\mathbf{X},t\right)$};

  %\draw[col3in,line width=2pt,-triangle 45] (2.5,0) -- (3.75,0);
  \hspace{-0.3cm}
  \draw[col3in,line width=2pt] (2.7,0) -- (3.75,0);
  \draw [rounded corners,color=col3in, fill=col3in] (4.1,0.) -- (3.5,-0.2)--(3.5,0.2)--cycle;
\end{scope}

\begin{scope}[xshift=-9cm, yshift=5cm] (theta)
\node[rectangle,draw=col2out!100,fill=col2in!10,inner sep=10pt, inner ysep=20pt, minimum width=4cm,minimum height=3cm,text justified,text=black] (mabox)
{ };
% le titre placé grace au label du noeud
\node[rectangle,draw=col2out!100,rounded corners,thick,fill=col2in!50,text=black]at (mabox.north)
{Prior Knowledge $\pi\left(\mathbf{X}\right)$};

  \begin{scope}[xshift=0.1cm, yshift=-0.65cm]
    \draw[<->] (-1,1.5) |- (1.5,0);
     \node[text width=2cm,text=black] at (0.9,-0.5) (x axis) {$ \mathbf{X}$};
     \node[text width=2cm,text=black] at (-0.9,0.65) (y axis) {$ \mathbf{\pi\left(\mathbf{X}\right)}$};
    \draw[-, color=green!30!black, thick]
      plot[id=xb, samples=2000 ,domain=-1.5:1.5]
      function{0.1*(exp(-(x*x-0.5)/0.2))};
  \end{scope}
  %\draw[col2in,line width=2pt,-triangle 45] (2,0) -- (3.5,0);
  \hspace{-0.25cm}
  \draw[col2in,line width=2pt] (2.25,0) -- (3.4,0);
  \draw [rounded corners,color=col2in, fill=col2in] (3.9,0.)--(3.35,-0.2)--(3.35,0.2)--cycle;
\end{scope}

\begin{scope}[xshift=-2cm, yshift=1cm] (theta)
% le titre placé grace au label du noeud
\node[rectangle,draw=col1out!50,rounded corners,thick,fill=col1in!50, minimum width=7cm,minimum height=1cm,text=black]
{\large{Model Calibration Algorithm}};
% \draw[col1in,line width=2pt,-triangle 45] (-3.5,0.5)--(-3.5,2.5);
\end{scope}

\begin{scope}[xshift=-3.5cm, yshift=8.cm] (theta)
% le titre placé grace au label du noeud
\node[rectangle,draw=col3out!50,rounded corners,thick,fill=col3in!50, minimum width=8cm,minimum height=1cm,text=black]
{\large{Direct Model Evaluation}};
 %\draw[col3out,line width=2pt,-triangle 45] (2.25,0)--(3.25,0);
 \draw[col3out,line width=2pt] (2.5,0)--(3.5,0);
 \draw [rounded corners,color=col3out, fill=col3out] (3.75,0.)--(3.25,-0.2)--(3.25,0.2)--cycle;
\end{scope}

\begin{scope}[xshift=2.5cm, yshift=5cm] (theta)
\node[rectangle,draw=col3out!100,fill=col3in!10,inner sep=10pt, inner ysep=20pt, minimum width=3.5cm,minimum height=3cm,text justified,text=black] (mabox)
{ };
% le titre placé grace au label du noeud
\node[rectangle,draw=col3out!100,rounded corners,thick,fill=col3in!50,text=black, minimum width=3cm]at (mabox.north)
{Result $\pi\left(\mathbf{O}\right)$};
\draw[col3in,line width=2pt] (-0.15,-1.5) -- (-0.15,-4.2);
\begin{scope}[xshift=-0.5cm, yshift=-0.9cm]
    \draw[<->] (0,2) |- (2,0);
     \node[text width=2cm,text=black] at (1.75,-0.4) (x axis) {$ \mathbf{O}$};
     \node[text width=2cm,text=black] at (0.1,0.75) (y axis) {$ \pi\left(\textbf{O}\right)$};
    \draw[-, color=green!30!black, thick]
      plot[id=xv, samples=2000 ,domain=0:2]
      function{0.75*x*(exp(-(x*x-0.5)/0.2))};
\end{scope}
\end{scope}
 
 \begin{scope}
 \node[rectangle, rounded corners, draw=col2out!50, fill=col2in!50, minimum height=1cm,text=black] at (2.5,-1.) (Obs){ Observations $\left(\mathbf{Y}\right)$};
 %\draw[col2in,line width=2pt,-triangle 45] (2.35,-0.5) -- (2.35,0.9)--(1.5,0.9);
 \draw[rounded corners,col2in,line width=2pt] (2.35,-0.5) -- (2.35,0.9)--(1.7,0.9);
 \draw [rounded corners,color=col2in, fill=col2in] (2.,0.7) -- (2.,1.1)--(1.5,0.9)--cycle;
  \end{scope}
  
  \begin{scope}[xshift=-9cm, yshift=1cm] (theta)
\node[rectangle,draw=col1out!100,fill=col1in!10,inner sep=10pt, inner ysep=20pt, minimum width=4.5cm,minimum height=3cm,text justified,text=black] (mabox)
{ };
% le titre placé grace au label du noeud
\node[rectangle,draw=col1out!100,rounded corners,minimum width=4cm,thick,fill=col1in!50,text=black]at (mabox.north)
{Posterior Knowledge $\pi\left(\mathbf{X}|\mathbf{Y}\right)$};

  \begin{scope}[xshift=0.1cm, yshift=-0.65cm, minimum width=4cm]
    \draw[<->] (-1,1.5) |- (1.5,0);
     \node[text width=2cm,text=black] at (0.9,-0.5) (x axis) {$ \mathbf{X}$};
     \node[text width=2cm,text=black] at (-1.35,0.65) (y axis) {$ \mathbf{\pi\left(\mathbf{X}|\mathbf{Y}\right)}$};
    \draw[-, color=green!30!black, thick]
      plot[id=xp, samples=2000 ,domain=-1.5:1.5]
      function{0.1*(exp(-(x*x-0.5)/0.18))};
  \end{scope}
  \hspace{0.2cm}
  %\draw[col1in,line width=2pt,triangle 45-] (2,0) -- (3.5,0);
  \draw[col1in,line width=2pt] (2.1,0) -- (3.3,0);
  \draw [rounded corners,color=col1in, fill=col1in] (2.0,0)--(2.5,0.2)--(2.5,-0.2)--cycle;
\end{scope}

\end{tikzpicture}
\end{adjustbox}
\caption{Model calibration under uncertainty using observations framework}
\label{fng:model_cali}
\end{figure}

Finally, the maximum \textit{a posteriori} is equivalent to the optimal search for the control vector minimizing the objective function $J\left({\bf{X}}\right)=\frac{1}{2}\left[{\bf{X}}-{\bf{X}}_0\right]\mathcal{B}^{-1}\left[{\bf{X}}-{\bf{X}}_0\right]'+\frac{1}{2}\left[{\bf{Y}}-G\left({\bf{X}}\right)\right]\mathcal{R}^{-1}\left[{\bf{Y}}-G\left({\bf{X}}\right)\right]'$ (where $\mathcal{B}$ is the \textit{prior} covariance matrix). This is known as the traditional variational data assimilation cost function, called 3D-VAR \citep{Carrassi_2018}. 

Mathematical methods can be used to solve optimization problems. The former can vary significantly according to the form of the cost function (convex, quadratic, nonlinear, etc.), its regularity, and the dimension of the space. Many deterministic optimization methods are known as gradient descent methods, among which is the Broyden-Fletcher-Goldfarb-Shanno (BFGS) quasi-Newton method us\-ed in this work \citep{Morales_2011}. Using this constrained optimization method makes it possible to impose boundaries during the research process of the model parameters, guaranteeing their physical values. Because the inverse problem is often ill-posed and unstable with available data corresponding to more than one solution, small changes in model results can lead to very different estimates for the input (calibration) parameters. These problems are related to the issue of “parameter identifiability” \citep{Navon_1998}. Still, the chosen optimization method involves computing the adjoint of the observation operator $G$ (or the partial derivatives of the operator with respect to its input parameters). In this work, the partial derivatives are approximated using a classical finite difference method. This is a simple solution, numerically sensitive and computationally time-consuming, but the observation operator can be written to make use of parallel computing, thus providing a fast automatic calibration algorithm relevant to real-world applications.

As the inverse problem is defined, a relevant question is what are the effects of the modeling calibration parameters ${\bf{X}}$ on the simulated state variables $O\left(t\right)=M\left({\bf{X}},t\right)$ which are compared to the observations? This question can be addressed by multivariate sensitivity analysis.

\subsection{Multivariate Sensitivity Analysis}
\label{subsec:sensitivity}

The sensivity analysis aims at quantifying the relative importance of each input model parameter. The variance-based methods aim at decomposing the variance of the output to quantify the participation of variables considered as independent. Conventional approaches to Global Sensitivity Analysis (GSA) compute sensitivity indices called Sobol' indices and assume that the output variable is scalar. The definition of Sobol' indices is a result of the ANOVA (ANalysis Of VAriance) variance decomposition \cite{Sobol_2001,Saltelli_2002}. However, as reported by many authors \citep{Campbell_2006,Lamboni_2014,Garcia-Cabrejo_20014}, when this approach is applied to each variable of a functional output, it leads to a high degree of redundancy because of the strong relationship between responses from one time step to another. Moreover, it also misses important features of the output dynamic because many features cannot be efficiently detected through single-time measurements. Thus, the methodology used in this work to compute GSA is a multivariate sensitivity analysis computing ``Generalized Sensitivity Indices'' \cite{Lamboni_2014} which are similar to ``Aggregated Sensitivity Indices'' \cite{Gamboa_2013}, as demonstrated in \cite{Garcia-Cabrejo_20014,Marrel_2017}.

\subsubsection{Analysis of Covariance}

To perform sensitivity analysis, some parameters and input variables are identified as uncertain, while the others are fixed at given nominal values. \telemacdd denoted $M$ thereafter ensures the relationship between the vector of model inputs ${\bf{X}}$ and the output quantity of interest $O\left(t\right)$ whereby $O\left(t\right) = M\left({\bf{X}},t\right)$ (``Direct Model Evaluation'', Figure \ref{fng:model_cali}). Since \telemacdd is a discrete time dynamic model, for one interest point, the model output ${\bf{O}}$ is composed of scalar output given at $t\in \mathbb{T}=\left[1,..,T\right]$ such as ${\bf{O}}=\left(O_1,...,O_T\right)$ (with $O_i=O\left(t_i\right)=M\left({\bf{X}},t_i\right)$). The covariance of the model vector output ${\bf{O}}=\left(O_1,...,O_T\right)$ can be partitioned into summands of increasing dimensions (single, pairs, triplets and so on) of input variables, as shown in Eq. \ref{eq:decomp_cov}.

\begin{equation}
Cov\left(O_1,...,O_T\right)=\sum_{i=1}^{p}Cov_i\left(O_1,...,O_T\right)+\sum_{1\leq i\le j\leq p}^{P}Cov_{i,j}\left(O_1,...,O_T\right)+...+Cov_{1,...,P}\left(O_1,...,O_T\right)
\label{eq:decomp_cov}
\end{equation}

From this equation, some sensitivity indices of multivariate case can be defined by projection of the covariance matrix into scalar using the trace operator. \citet{Gamboa_2013} present the following aggregated indices:

\begin{equation}
    \left\{
    \begin{array}{ll}
        S1_i\left(O_1,...,O_T\right)=&\frac{Tr\left[Cov_i\left(O_1,...,O_T\right)\right]}{Tr\left[Cov\left(O_1,...,O_T\right)\right]}\\
        ST_i\left(O_1,...,O_T\right)=&\frac{Tr\left[Cov_i\left(O_1,...,O_T\right)\right]+\sum_{1\leq i\le j\leq p}^{p}Tr\left[Cov_{i,j}\left(O_1,...,O_T\right)\right]+...+Tr\left[Cov_{1,...,p}\left(O_1,...,O_T\right)\right]}{Tr\left[Cov\left(O_1,...,O_T\right)\right]}
    \end{array}
\right.
\label{eq:indice_multi}
\end{equation}

\subsubsection{Computation of the multivariate sensitivity indices}

To handle dynamic system behavior under parameter uncertainty, a sample set of configurations is generated using Latin Hypercube Sampling (LHS). In the sampling procedure, the parameter uncertainties are taken from uniform distribution whose limits are defined by the parameter bounds (see Section \ref{sub:range_var}). For a fixed number of desired configurations $n$, $n^p$ LHSs are feasible. It is therefore possible to select the plan optimizing a uniformity criterion (discrepancy, \ldots) or a statistical criterion (entropy, \ldots). In this study, the uniformity criterion relying on $L^2$-discrepancies was considered and optimized, as described in \cite{Damblin_2013}. In fact, according to these autors, the $L^2$-discrepancy optimized LHS have shown some good properties to 2D projections in high dimension. The full set of output dynamics over the complete Latin Hypercube Sampling gives a matrix of size $n\times T$.

\begin{equation}
\mathcal{O} = \left({\bf{O}}_1,...,{\bf{O}}_T\right)=  \begin{pmatrix}
  o_{1,1} & o_{1,2} & \cdots & o_{1,T} \\
  o_{2,1} & o_{2,2} & \cdots & o_{2,T} \\
  \vdots  & \vdots  & \ddots & \vdots  \\
  o_{n,1} & o_{n,2} & \cdots & o_{n,T} 
 \end{pmatrix}
\label{eq:reali}
\end{equation}

Each row of the matrix $\mathcal{O}$ corresponds to a temporal trajectory for a fixed scenario of uncertain input parameter. As proposed by \citet{Lamboni_2014}, a Principal Component Analysis (PCA) can be carried out on the matrix $\mathcal{O}_c$ which is the matrix obtained by centering each column of $\mathcal{O}$. The PCA decomposition consists in a small number determination of variable $\left({\bf{H}}_1,...,{\bf{H}}_T\right)$ to transcribe the maximum information, in the sense of variance, contained in the initial variables $\mathcal{O}_c=\left({\bf{O}}_1^c,...,{\bf{O}}_T^c\right)$ (with ${\bf{O}}_i^c={\bf{O}}_i-\mathbb{E}\left({\bf{O}}_i\right)$, where $\mathbb{E}\left(.\right)$ is the mean operator and ${\bf{O}}_i$ the discrete form of $O_i$) in order to remove redundancies with the least possible loss of information. The variables ${\bf{H}}_i$ are constructed to be uncorrelated from each other, based on an iterative process that adds complementary information not contained in the previously constructed variables. This methodology can be summarized by resolution of the system described by Eq. \ref{eq:acp_pb_syst}.

\begin{equation}
    \left\{
    \begin{array}{ll}
        {\bf{H}}_i=\mathcal{O}_c{\bf{w}}_i, i=1,...,T\\
        \text{with }Var\left({\bf{H}}_1\right)\geq Var\left({\bf{H}}_2\right)...Var\left({\bf{H}}_T\right)>0\\
        \text{ under the constraint} \left\lVert {\bf{w}}_i \right\rVert^2 = 1
    \end{array}
\right.
\label{eq:acp_pb_syst}
\end{equation}

In practice, $\left({\bf{w}}_1,...,{\bf{w}}_T\right)$ is a set of normalised and mutually orthogonal eigenvectors associated with eigenvalues $\left(\lambda_1,...,\lambda_T\right)$ of the covariance matrix $\Sigma=\frac{1}{n}\mathbb{E}\left[\mathcal{O}_c^T\mathcal{O}_c\right]$. Once the eigenvectors resulting from the diagonalization of $\Sigma$ are determined, the matrix $\mathcal{O}_c$ can be decomposed as written in Eq. \ref{eq:decompo_Snap}.

\begin{equation}
\mathcal{O}_c =\sum_{i=1}^{T}{\bf{H}}_i{\bf{w}}_i'
\label{eq:decompo_Snap}
\end{equation}

${\bf{H}}_i$ variables are uncorrelated, and the principal component matrix $\mathcal{H} =  \left({\bf{H}}_1,...,{\bf{H}}_T\right)$ has the same total inertia as $\mathcal{O}$, but is mostly concentrated in its first columns. Thus, the expression of covariance decomposition can be expressed as a sum of variance of ${\bf{H}}_i$ on which the Sobol' decomposition into summands of increasing dimension can be carried out (Eq. \ref{eq:decomp_ANOVA}).

\begin{align}
   \left\{
    \begin{array}{ll}
        Tr\left[Cov\left(O_1,...,O_m\right)\right] =\sum_{k=1}^T\left(\sum_{i=1}^pV_{\left(i\right),k}+\sum_{1\leq i\le j\leq p}^pV_{\left(i,j\right),k}+...+V_{\left(1,...p\right),k}\right)\\
        \text{with }
        V_{\left(\epsilon\right),k}=Var\left[\mathbb{E}\left({\bf{H}}_k|{\bf{X}}_{\epsilon}\right)\right]-\sum_{w,w\in\epsilon}V_{w,k}\\
\end{array}
  \right.
\label{eq:decomp_ANOVA}
\end{align}

where $Var\left[\mathbb{E}\left({\bf{H}}_k|{\bf{X}}_{\epsilon}\right)\right]$ is the variance of the conditional expectation of ${\bf{H}}_k$ to the input variable ${\bf{X}}_\epsilon$

Thus, based on an identification process between the formulation that is described in Eq. \ref{eq:indice_multi} and Eq. \ref{eq:decomp_ANOVA}, the expression of multivariate sensitivity indices can be given, as described in Eq. \ref{eq:indice_multi_acp}.

\begin{equation}
    \left\{
    \begin{array}{ll}
        S1_i\left(O_1,...,O_m\right)=& \frac{\sum_{k=1}^T S_{\left(i\right),k}^{ob}*\lambda_k}{\sum_{k=1}^T \lambda_k}\\
        ST_i\left(O_1,...,O_m\right)=&\frac{\sum_{k=1}^T S^{ob}T_{\left(i\right),k}*\lambda_k}{\sum_{k=1}^T \lambda_k}
    \end{array}
\right.
\label{eq:indice_multi_acp}
\end{equation}

where $S_{\left(\epsilon\right),k}^{ob}$ denotes Sobol' sensitivity indices such as $S_{\left(\epsilon\right),k}^{ob}=\frac{V_{\left(\epsilon\right),k}}{Var\left({\bf{H}}_k\right)}$ and $S^{ob}T_{\left(i\right),k}$ represents the total Sobol' sensitivity index such as $S^{ob}T_{\left(i\right),k}=\frac{\sum_{i\in\epsilon}V_{\left(\epsilon\right),k}}{Var\left({\bf{H}}_k\right)}$.

As expressed by Eq. \ref{eq:indice_multi_acp}, the multivariate sensitivity analysis indices involve Sobol' index computation. The next paragraph is dedicated to the computation process of this mathematical expression based on a combined PCA and Polynomial Chaos Expansion (PCE) emulator.

\subsubsection{Reduced Order Model PCA-PCE}
\label{sub:rom}

As already mentioned, Python is a language that can be used in many scientific contexts and can be adapted to any type of use based on dedicated libraries. The multivariate sensitivity analysis explained in the previous section was carried out based on an open-source library for uncertainty treatment named ``OpenTURNS'', standing for ``Open source initiative to Treat Uncertainties, Risks’N Statistics'' (\url{www.openturns.org}) \cite{Baudin_2017}. A Design of Experiment (DoE) of size $1000$ ($n=1000$) is constructed for the PCA-PCE learning step. This number of model evaluation was determined based on a convergence study carried on the sensitivity analysis (see Section \ref{sub:gsa}). The MPI library was used for launching and managing the \telemacdd solver computations. At the end of the calculation, a \telemacdd result file was composed of $288$ time records corresponding to a physical variable record every $10$ minutes ($T=288$). In the present study, the discrete version of the PCA (singular value decomposition (SVD)) of the matrix $\mathcal{O}_c$ is considered. This means that ${\bf{H}}_i$ considered is expressed as ${\bf{H}}_i=\sqrt{n \lambda_i}{\bf{A}}_i$ with ${\bf{A}}_i$ orthonormal. In order to have an efficient sensitivity index computation in terms of computational cost, the $k$ main eigenvectors which explained more than $99.95$\% of the original variance are considered ($k\sim 10<<T$). It is important to note that, owing to truncation, the equivalence between ``Aggregated Sensitivity Indices'' \cite{Gamboa_2013} and ``Generalized Sensitivity indices'' \cite{Lamboni_2014} no longer exists. The Generalized Sensitivity indices represent an approximation of the Aggregated Sensitivity Indices. For each eigenvector, a learning sample of realizations of the decomposition coefficient ${\bf{H}}_i$ is available. As proposed by \citet{Garcia-Cabrejo_20014}, a polynomial Chaos Expansion can be used as learning step of each decomposition coefficient. Thus, the Sobol' indices needed in Eq. \ref{eq:indice_multi_acp} are obtained based on post-treatment of each mode PCE as proposed in \cite{Sudret2008}. Here, the construction of the PCE is carried out based on Least Angle Regression Stagewise method (LARS) in order to construct an adaptive sparse PCE. In this approach, a collection of possible PCEs, ordered by sparsity, is provided and an optimum can be chosen with an accuracy estimate. It was performed in this study using corrected leave-one-out error \citep{BLATMAN20112345}.

At this stage, a reduced order PCA-PCE model is produced for each observation station of the Gironde Estuary (Figure \ref{fig:estu_gir}). The validation of these emulators was carried out based on a validation sample generated with the same procedure (optimized LHS) as for the learning step. This sample was composed of $100$ validation points ($n_{val}=100$) and allows, after the \telemacdd unit computing, a temporal predictivity criterion defined by Eq. \ref{eq:Q2_temporal}.

\begin{equation}
  Q^2\left({t}\right)=1-\frac{\sum_{j=1}^{n_{val}}\left[\hat{M}_k\left({\bf{X}}_j,t\right)-M\left({\bf{X}}_j,t\right)\right]^2}{\sum_{j=1}^{n_{val}}\left[\hat{M}_k\left({\bf{X}}_j,t\right)-\frac{1}{n_{val}}\sum_{i=1}^{n_{val}}M\left({\bf{X}}_j,t\right)\right]^2}
  \label{eq:Q2_temporal}
\end{equation}

where $\hat{M}_k\left({\bf{X}}_j,t\right)$ is the estimated PCA-PCE evolution.

\begin{figure}[h!]
  \centering
  \begin{tabular}{cc}
     \subfloat[Verdon observation station\label{Verdon-rom-val}]{%
       \includegraphics[width=0.5\textwidth]{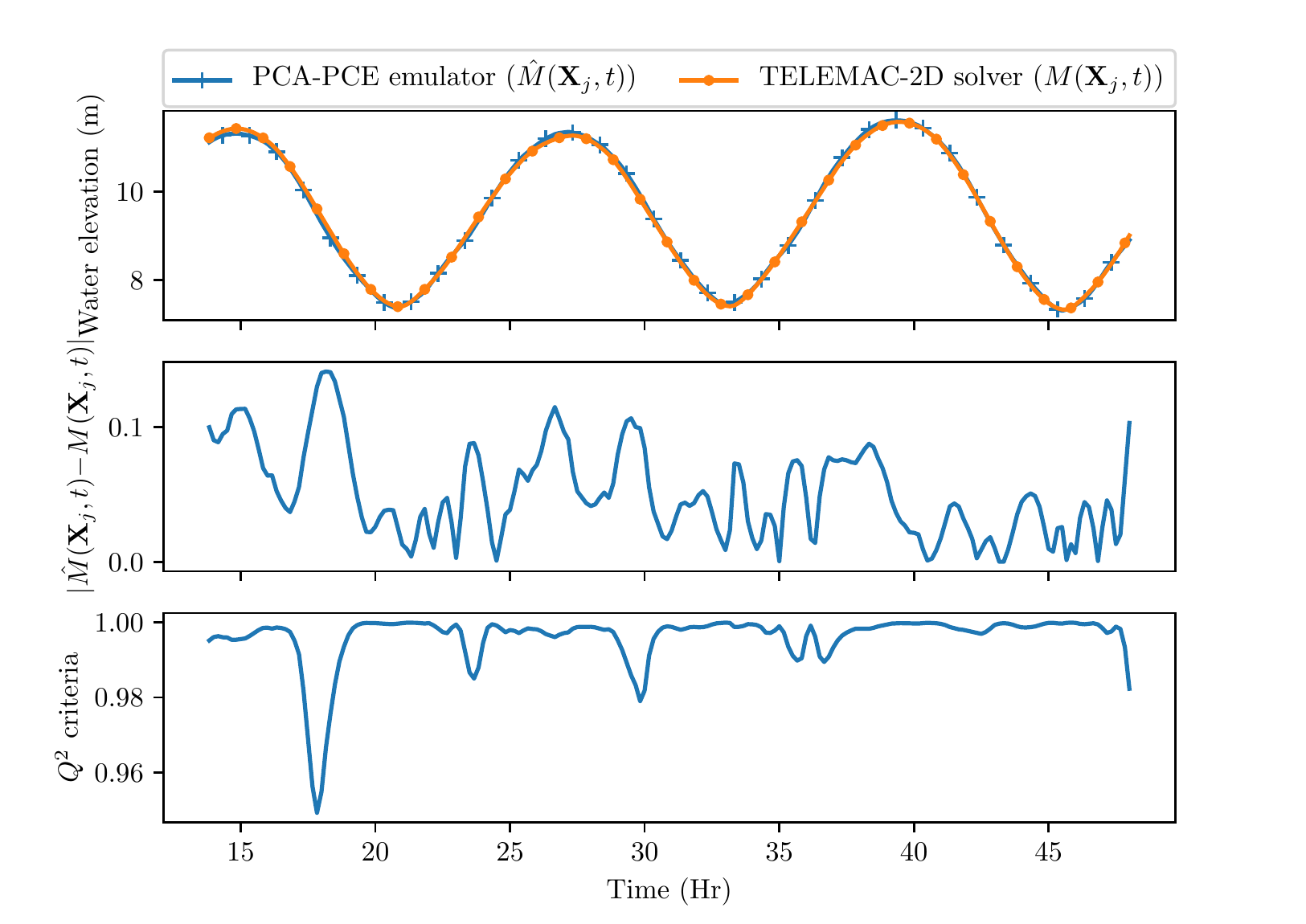}}&
     \subfloat[Bordeaux observation station\label{Bordeaux-rom-val}]{%
       \includegraphics[width=0.5\textwidth]{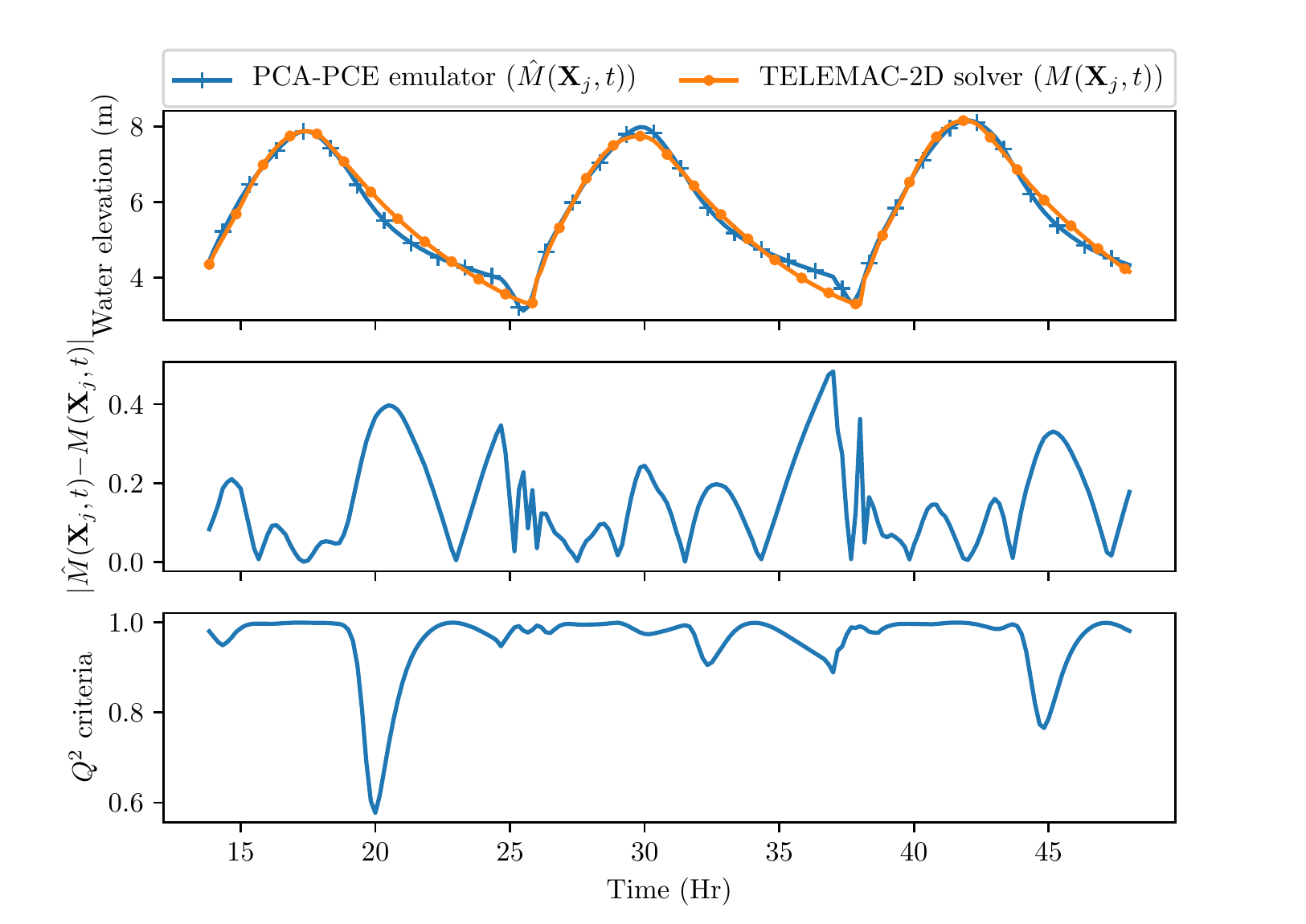}}
     \end{tabular}
    \caption{Validation of the reduced order PCA-PCE model}
    \label{fig:rom-val}
\end{figure}

Figure \ref{fig:rom-val} shows (from top to bottom) a comparison on a validation configuration between an estimated result from the reduced order PCA-PCE model and the \telemacdd computation. The difference between these two water elevations is then given, and finally, the predictivity criterion is presented. 

The reduced order model is validated according to the performance displayed in Figure \ref{fig:rom-val}. Indeed, the results provided by the emulator and the \telemacdd solver are fairly close. The deviation between these results varies respectively from  $10$ to $20$ centimeters at Verdon and Bordeaux observation stations. This is confirmed by the average predictivity criterion close to $1$ for Verdon and Bordeaux during the study period. A coefficient close to 1 shows a good fit between the validation database and the result estimated by the reduced order model. However, a less satisfactory performance of $Q^2$ criteria can be noticed at the Bordeaux observation station, compared to the Verdon station. The confluence of the Dordogne and Garonne rivers occurs just downstream from Bordeaux and the influence of the hydrological forcing of these two rivers is not studied here. Consequently, the upstream fluvial discharges are not considered as parameters in construction of the PCA-PCE emulator. The missing interactions can explain the observed performance gap in the predictivity criteria.

To conclude, the reduced PCA-PCE model shows good agreement with the \telemacdd computation. The high score of predictivity criteria allows confidence in the sensitivity results post-treated from the reduced order model.

\subsubsection{Multivariate sensitivity analysis results}

A reduced order model based on PCA-PCE is used to identify influential input parameters. As reported by \citet{Pianosi_2016}, when applying sampling-based sensitivity analysis, sensitivity indices are not computed exactly but they are approximated from the available samples. The robustness and convergence of such sensitivity estimates should therefore be assessed. Thus, three elements are of interest in the following analysis: (i) the convergence of sensitivity indices, (ii) the robustness of the estimates and (iii) relevant sensitivity analysis visualization.

\paragraph{Convergence:\newline}

\label{sub:gsa}

\begin{figure}[ht!]
\centering
\begin{tabular}{cc}
\subfloat[First order\label{GSI_FO}]{%
\includegraphics[width=0.42\textwidth]{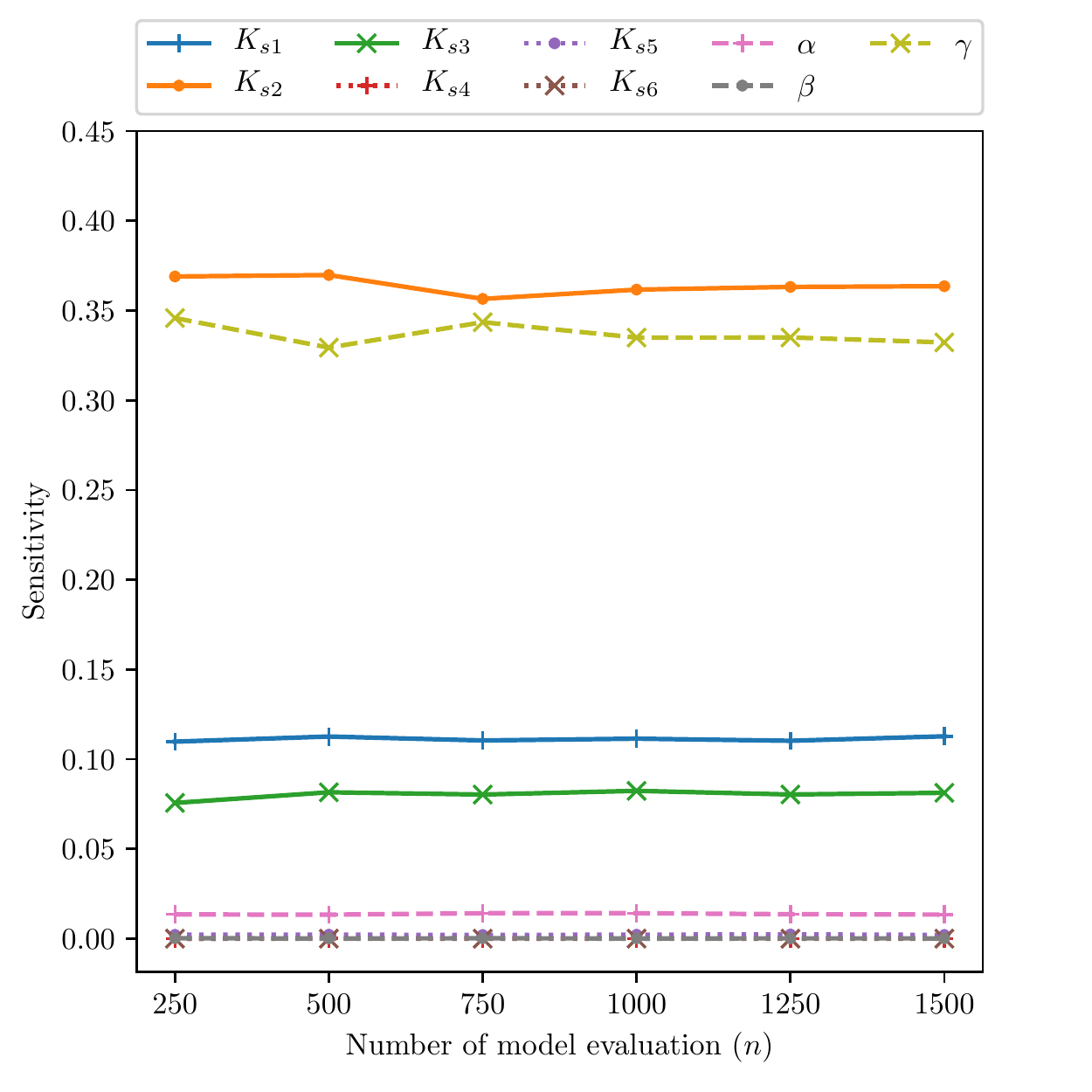}}&
\subfloat[Total order\label{GSI_TO}]{%
\includegraphics[width=0.42\textwidth]{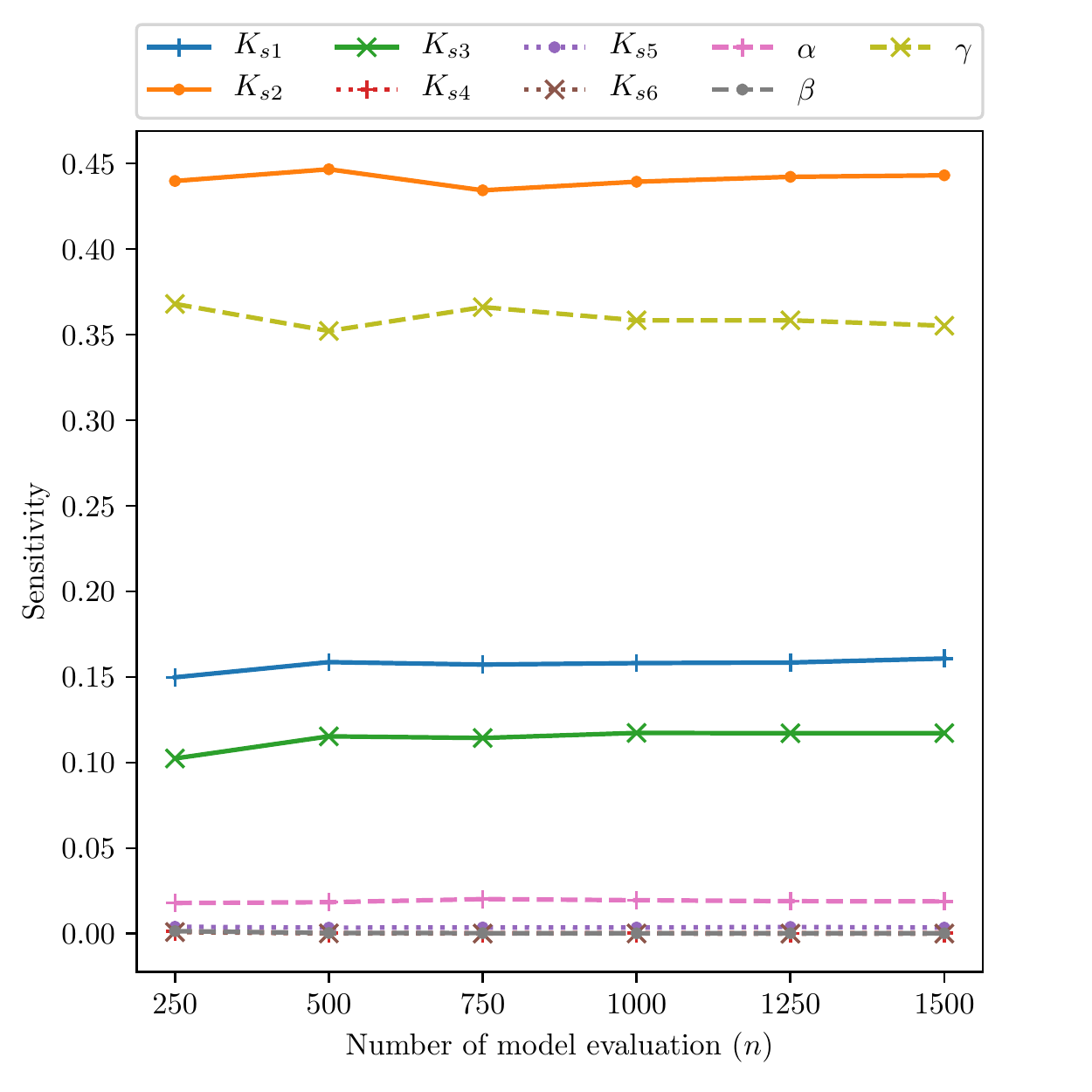}}
\end{tabular}
\caption{Generalized Sensitivity Indices estimated using an increasing sample size at Bordeaux observation station}
\label{fig:gsi_bordeaux}
\end{figure}

To handle this issue, the convergence rates of Generalized Sensitivity Indices are assessed as the sample size increases. Figure \ref{fig:gsi_bordeaux} presents the evolution of sensitivity indices obtained at Bordeaux observation station.

As displayed on Figure \ref{fig:gsi_bordeaux}, the number of samples needed to reach stable sensitivity estimates can vary from one input factor to another. However, from a sample size of $1000$, the sensitivity indices are stabilized. Thus, this number of model evaluation is considered in the following investigations.

\paragraph{Robustness:\newline}

A robustness analysis is carried out in order to evaluate the sensitivity of the estimates to the Design of Experiment. First, the robustness of the sensitivity indices is analysed through confidence intervals. They are obtained by repeating $30$ times the methodology described above (see Section \ref{sub:rom}). Table \ref{Tab:GSI_bordeau} presents the result of these confidence intervals, which have finally required $N = 30\times n = 3\times 10^4$ model calls.

\begin{longtable}{p{.13\textwidth} |llll}
\hline
\multirow{2}{*}{\diagbox{Inputs}{\textbf{GSI}}} & \multicolumn{2}{c}{\textbf{Verdon}}&\multicolumn{2}{c}{\textbf{Bordeaux}}\\
& First Order & Total Order & First Order & Total Order\\
\hline
\hline
$K_{s1}$ &  $\left[0.123,0.131\right]$   & $\left[0.131,0.139\right]$  &$\left[0.108,0.114\right]$ & $\left[0.153,0.162\right]$\\
$K_{s2}$ &  $\left[0.00969,0.0103\right]$   & $\left[0.0148,0.0157\right]$  &$\left[0.352,0.367\right]$ & $\left[0.43,0.448\right]$\\
$K_{s3}$ &  $\left[1.03,2.98\right]\times 10^{-5}$   & $\left[4.37,7.93\right]\times 10^{-5}$  &$\left[0.0790,0.0827\right]$ & $\left[0.115,0.119\right]$\\
$K_{s4}$ &  $\left[0.,1.79\times 10^{-6}\right]$   & $\left[0.275,2.84\right]\times 10^{-5}$  &$\left[0.561,1.59\right]\times 10^{-4}$ & $\left[2.58,4.21\right]\times 10^{-4}$\\
$K_{s5}$ &  $\left[1.32,3.12\right]\times 10^{-5}$   & $\left[4.62,7.35\right]\times 10^{-5}$  &$\left[1.96,2.40\right]\times 10^{-3}$ & $\left[3.46,4.01\right]\times 10^{-3}$\\
$K_{s6}$ &  $\left[0.,1.46\times 10^{-7}\right]$   & $\left[0.241,1.41\right]\times 10^{-5}$  &$\left[0.,8.49\right]\times 10^{-5}$ & $\left[0.772,2.17\right]\times 10^{-4}$\\
$\alpha$ &  $\left[0.0452,0.0481\right]$   & $\left[0.0479,0.0509\right]$  &$\left[0.0131,0.0145\right]$ & $\left[0.0184,0.0203\right]$\\
$\beta$  &  $\left[4.35,5.15\right]\times 10^{-4}$   & $\left[4.55,5.40\right]\times 10^{-4}$  &$\left[1.25,2.37\right]\times 10^{-4}$ & $\left[2.45,4.16\right]\times 10^{-4}$\\
$\gamma$ &  $\left[0.802,0.813\right]$   & $\left[0.803,0.813\right]$  &$\left[0.326,0.352\right]$ & $\left[0.350,0.374\right]$\\
\hline
\hline
\caption{Min-max confidence intervals of Generalized Sensitivity Indices (GSI) obtained from $30$ repetitions of the computational process at Bordeaux and Verdon observation stations}
\label{Tab:GSI_bordeau}
\end{longtable}

As shown in Table \ref{Tab:GSI_bordeau}, the bounds of the min-max confidence intervals are relatively close and demonstrate the capacity of the reduced order PCA-PCE model, constructed from  a $L^2$-discrepancy optimized LHS sample, to produce precise estimates. An alternative sampling method based on low discrepancy sequence of Sobol is also considered. The obtained results are presented in Table \ref{Tab:lds_GSI_bordeau}. Most of the time, the sensitivity indices obtained with the low discrepancy sequence are included in the min-max confidence intervals obtained from $L^2$-discrepancy optimized LHS sample. When out of range, values stay close to the bounds. Consequently, the estimates can be considered satisfactory in terms of accuracy and robustness.

\paragraph{Results visualization:\newline}

\begin{figure}[h!]
  \centering
  \begin{tabular}{cc}
     \subfloat[Verdon station\label{Verdon-chord_graph}]{%
       \includegraphics[width=0.45\textwidth]{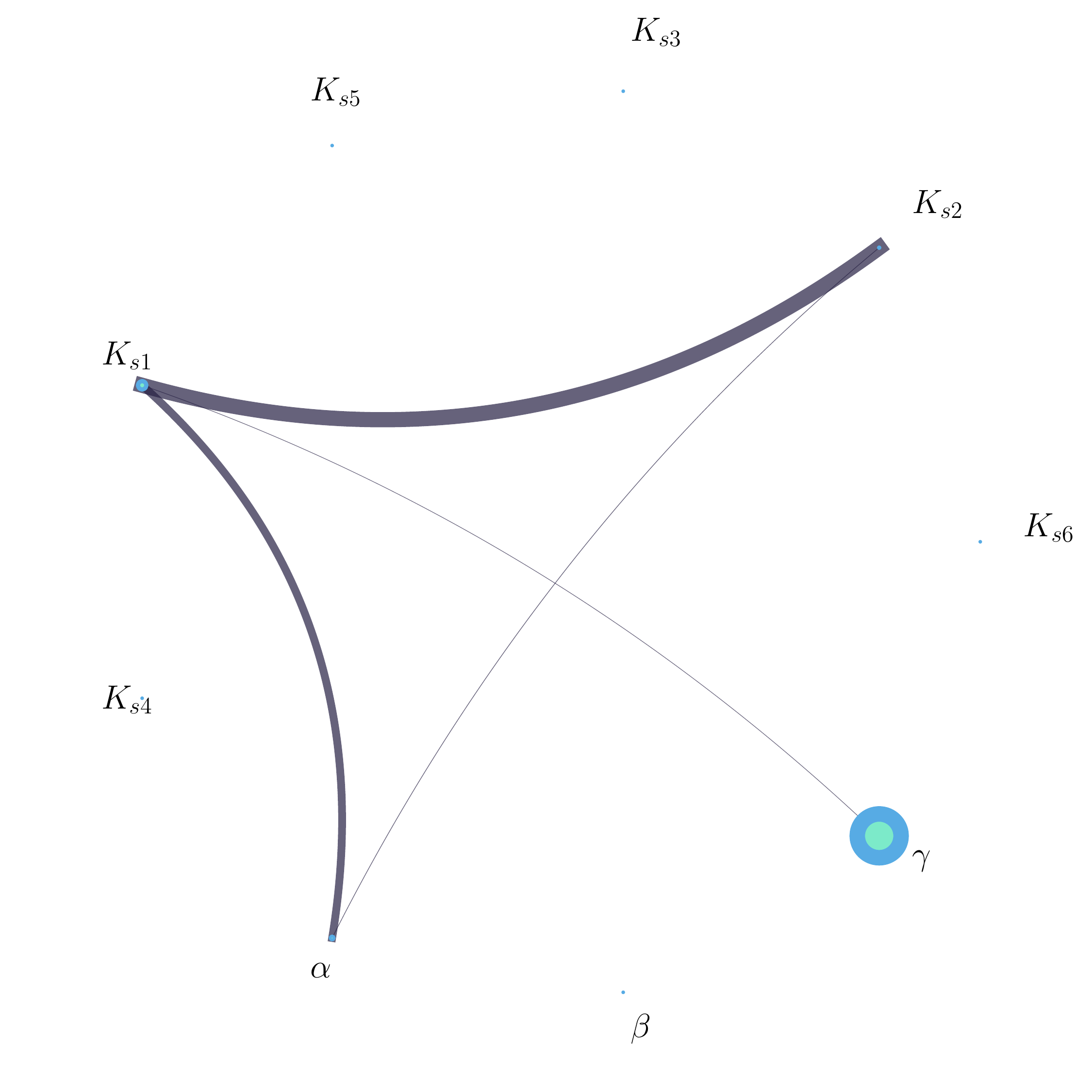}}&
     \subfloat[Lamena station\label{Lamena-Chord_graph}]{%
       \includegraphics[width=0.45\textwidth]{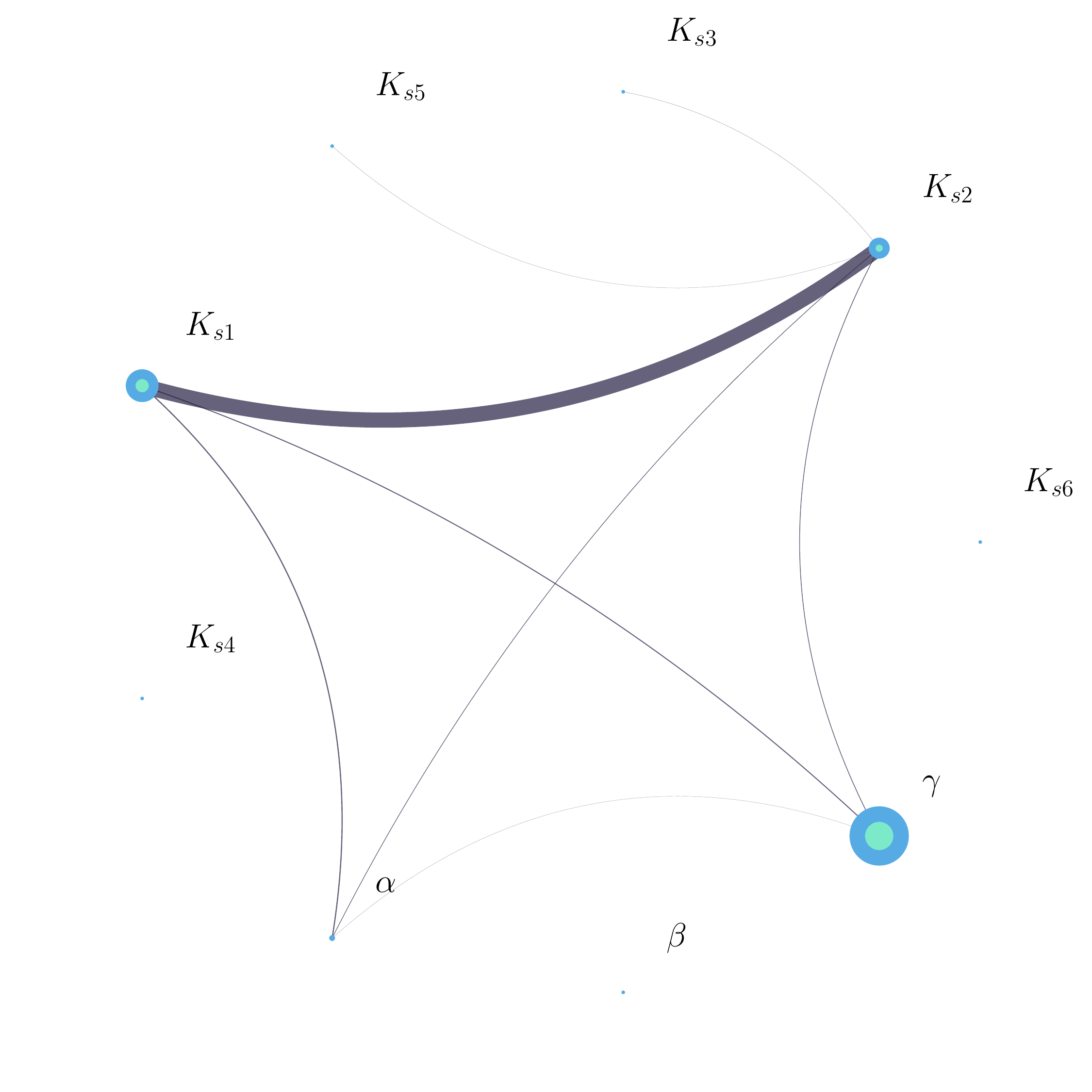}}\\
     \subfloat[Pauillac station\label{Pauillac-Chord_graph}]{%
       \includegraphics[width=0.45\textwidth]{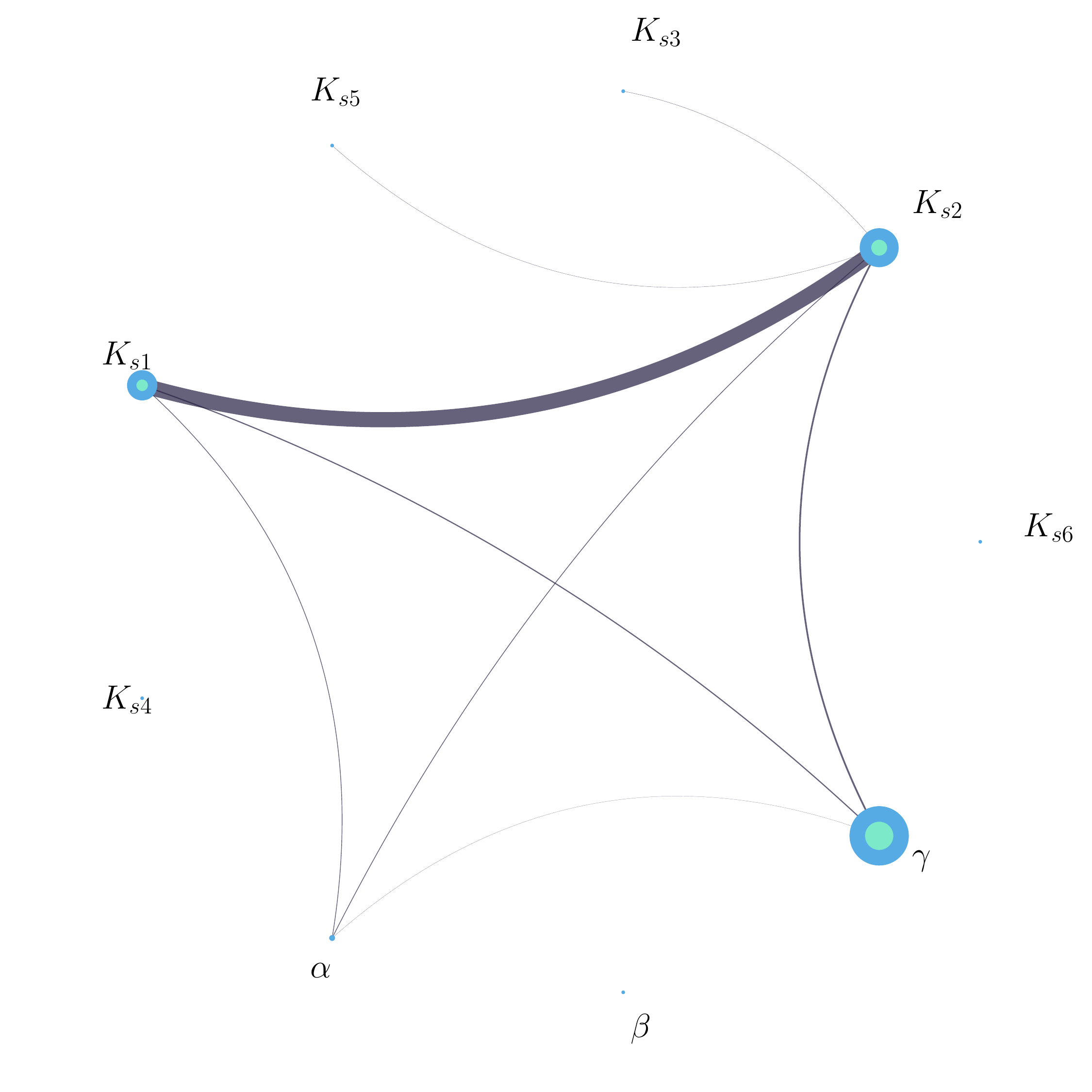}}&
     \subfloat[Bordeaux station\label{Bordeaux-Chord_graph}]{%
       \includegraphics[width=0.45\textwidth]{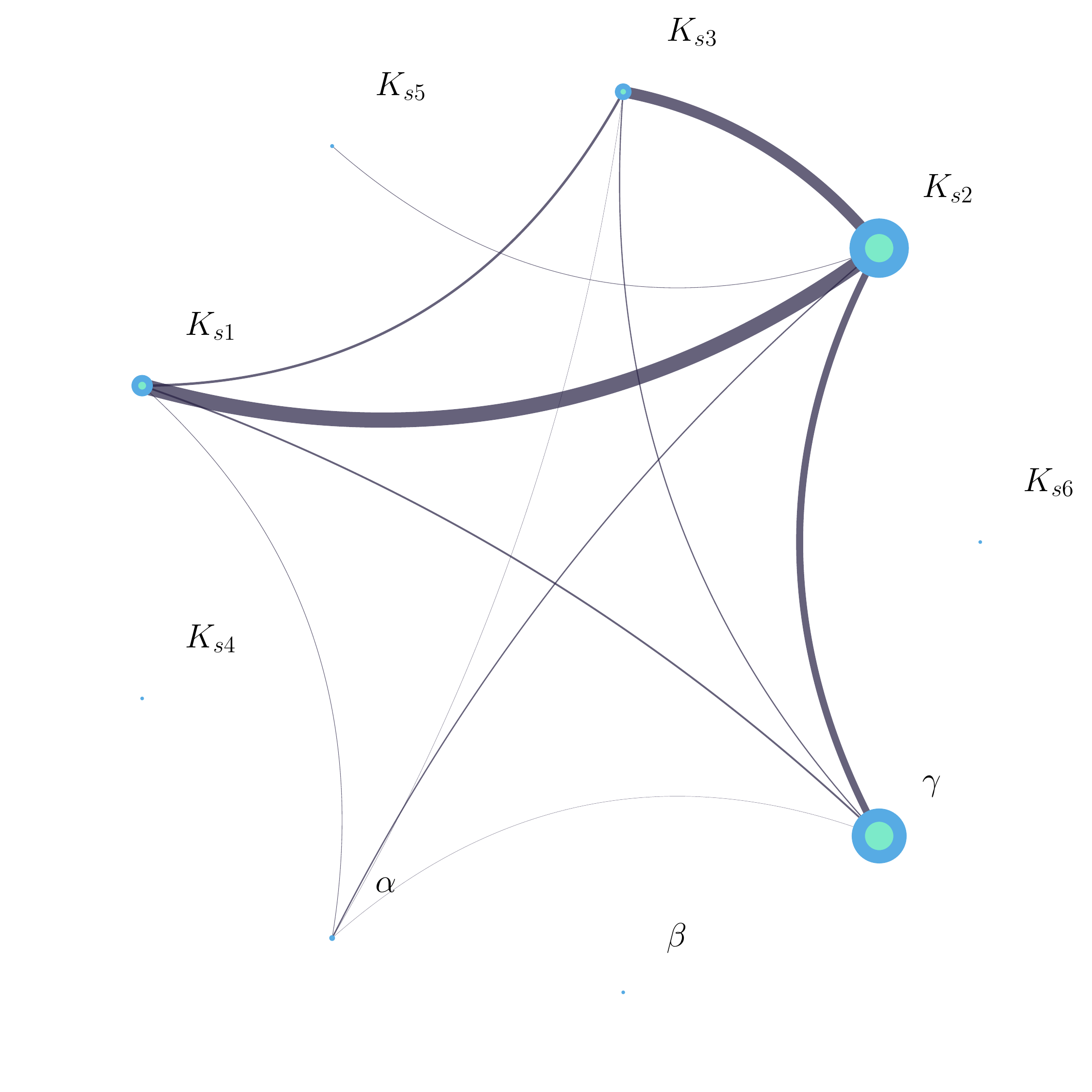}}\\
     \end{tabular}
    \caption{Chord graph of Generalized Sensitivity analysis}
    \label{fig:Chord_graph}
\end{figure}

Visualization can significantly improve the interpretation of the sensitivity analysis results. For this purpose, radial convergence diagrams, also called chord graphs, are used to simultaneously visualize some computed Generalized Sensitivity Indices (Figure \ref{fig:Chord_graph}). These diagrams plot the main effect of each input variable (first-order multivariate sensitivity analysis proportional to the size of the inner circle); its total influence, including interactions (proportional to the size of the outer circle), existence and extent of second-order effects (second-order multivariate sensitivity indices proportional to width) \citep{Butler_2014}.

\newpage

Figure \ref{fig:Chord_graph} displays sensitivity analysis estimates from one $L^2$-discrepancy optimized LHS sample used to compute the min-max confidence intervals presented in Table \ref{Tab:GSI_bordeau}. As shown in the Figure, the output variance at each observation station is mainly explained by the upstream friction coefficients (areas one and two of the Gironde Estuary model) and the sea water level. As expected, downstream from the estuary (Bordeaux station), the friction coefficient of area $3$ containing the Bordeaux station has a more significant influence. It is noticeable, from the visualization of variable interactions, that even if the tidal range variable contribution is not considerable, its interaction with other parameters should not be neglected. This enhances the utility of simultaneously visualized interactions with main and total effects of sensitivity analysis.

To conclude, the calibration problem initially composed of nine input variables can be reduced to five (the first three being friction area, sea level correction and tidal range coefficients) after considering the results of multivariate sensitivity analysis.

\pagebreak

\subsection{Calibration Results}

\begin{figure}[h!]
\centering
\begin{tabular}{cc}
\subfloat[Global cost function\label{J_cost_func}]{%
\includegraphics[width=0.45\textwidth]{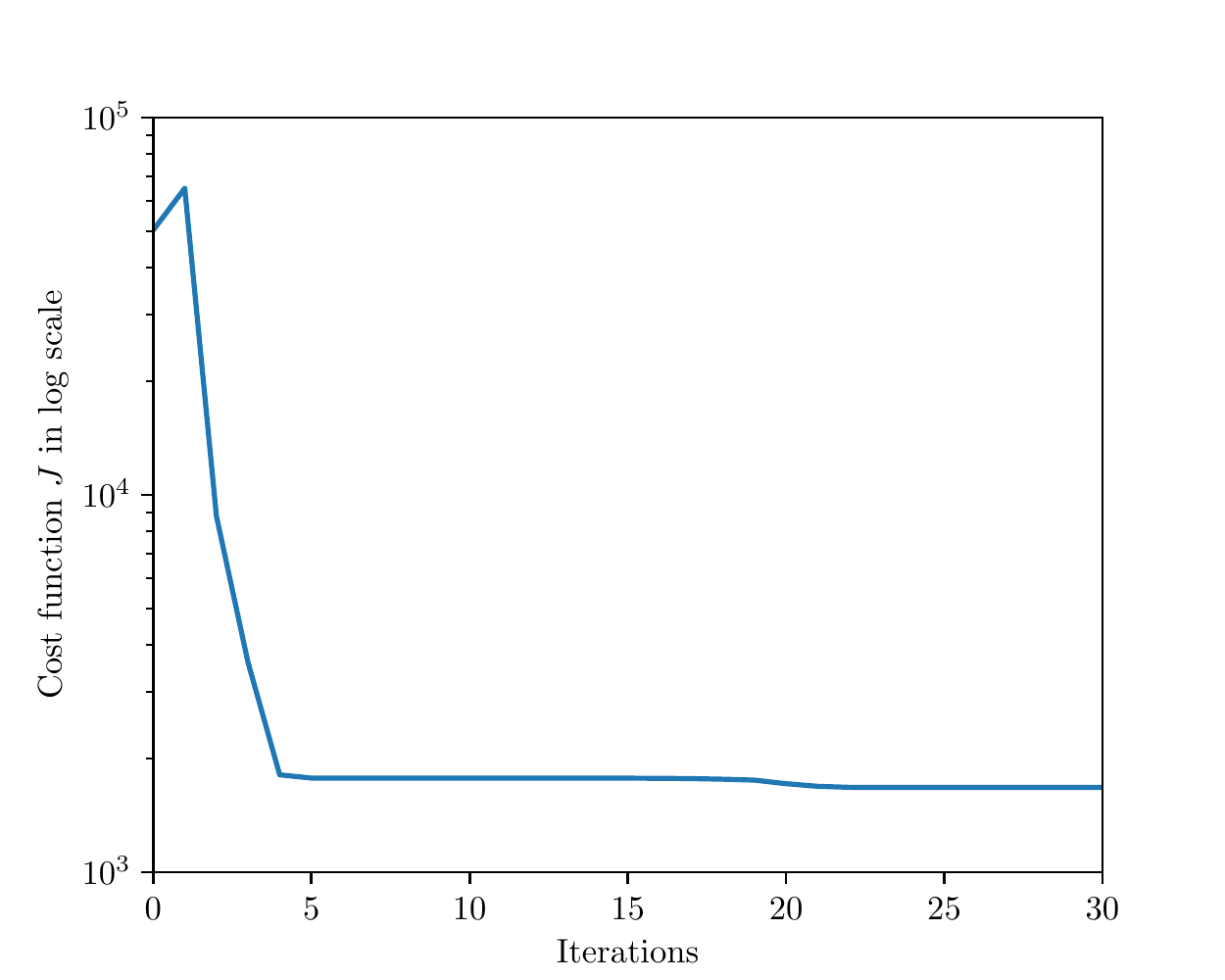}}&
\subfloat[Cost function of the prior and observation part\label{Cost_function_multi}]{%
\includegraphics[width=0.45\textwidth]{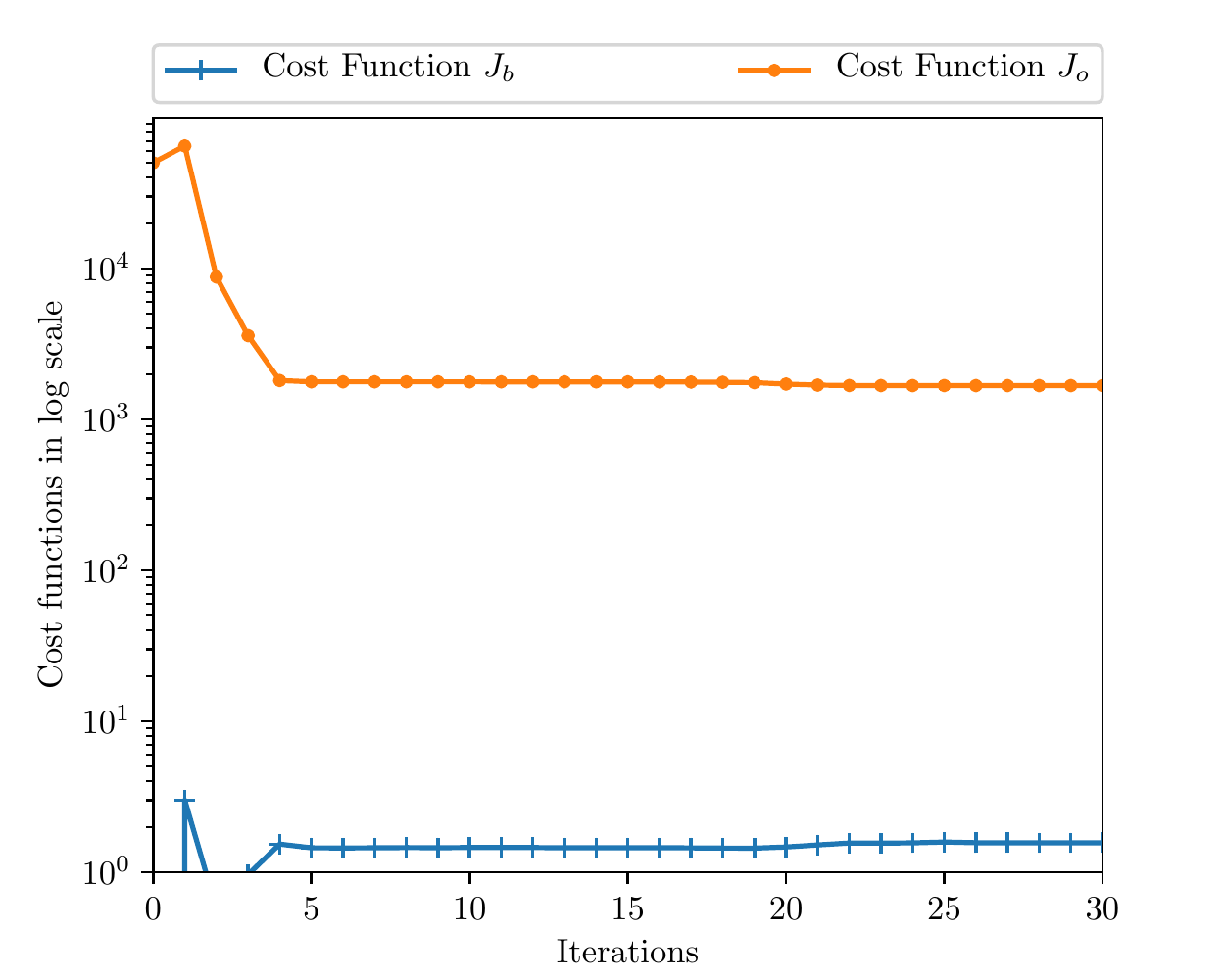}}
\end{tabular}
\caption{Value of cost functions according to number of algorithmic calibration iterations}
\label{fig:cost_func}
\end{figure}

\begin{figure}[h!]
\centering
\begin{tabular}{cc}
\subfloat[Friction coefficients\label{cf_estimated}]{%
\includegraphics[width=0.45\textwidth]{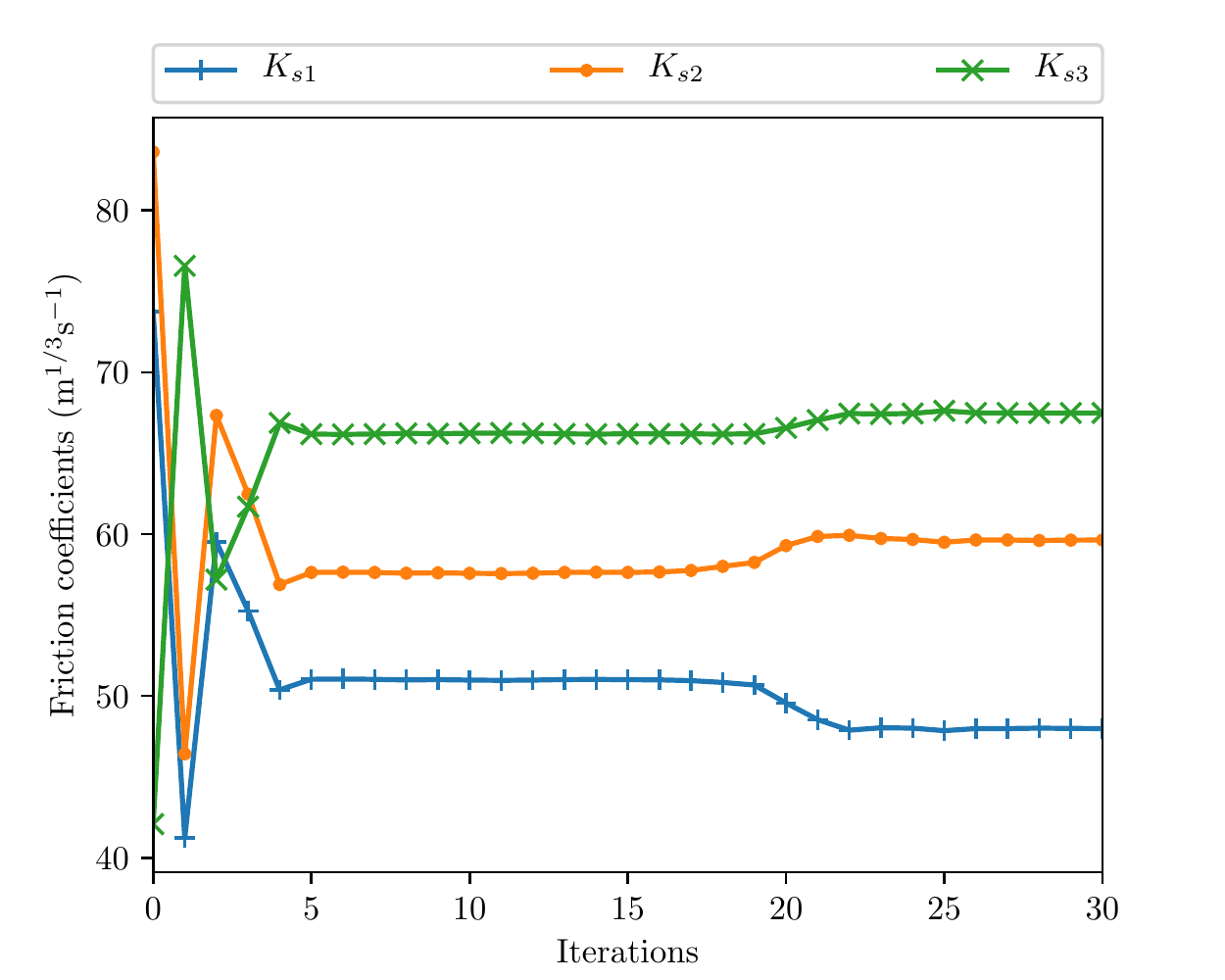}}&
\subfloat[Tidal parameters\label{tidal_estimated}]{%
\includegraphics[width=0.45\textwidth]{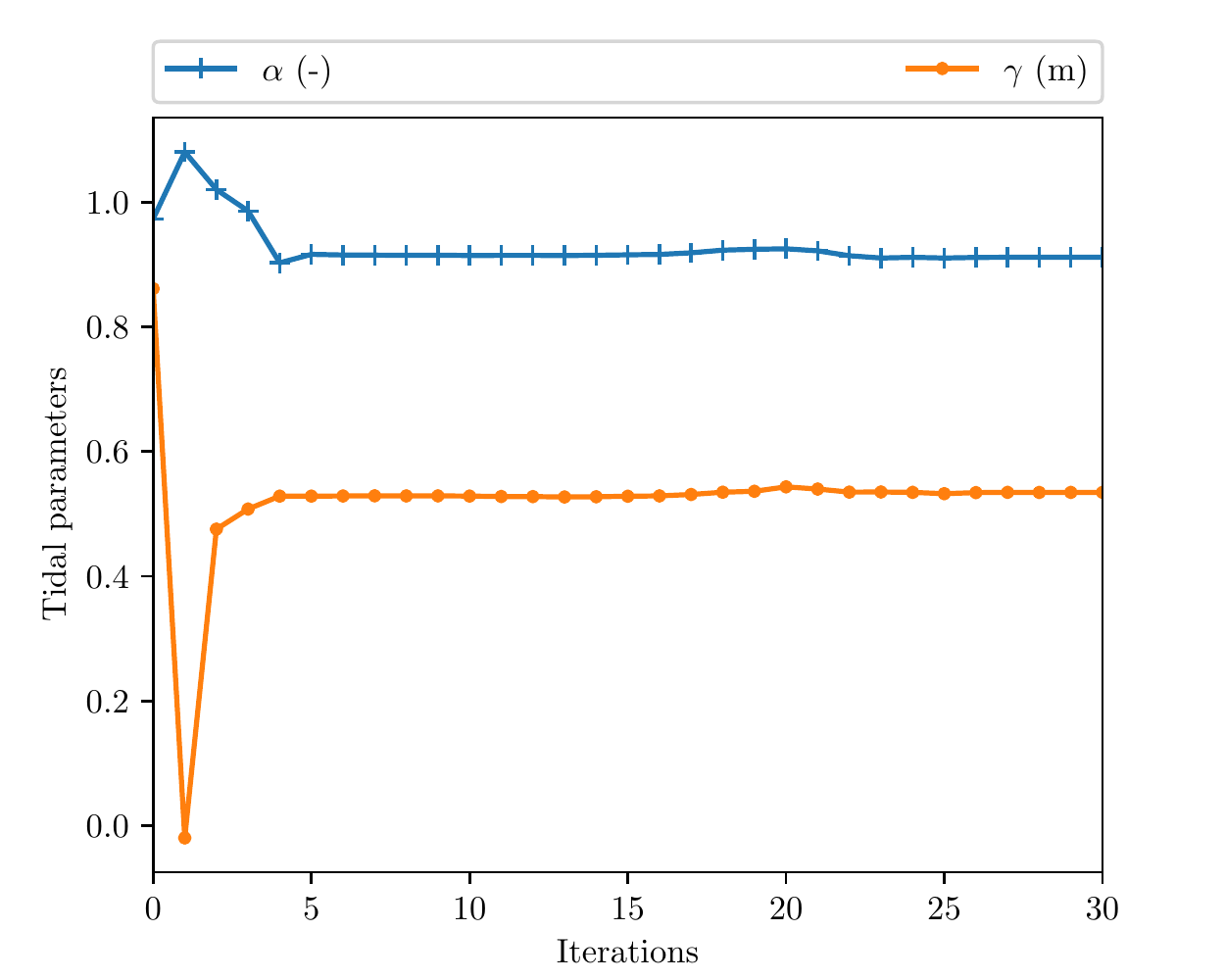}}
\end{tabular}
\caption{Value of the parameters according to number of algorithmic calibration iterations}
\label{fig:param_esti}
\end{figure}

The calibration algorithm (Figure \ref{fng:model_cali}) is performed by coupling \telemacdd and the data assimilation library ADAO in Python context through the component TelApy of the \telemacsystem. ADAO, A module for Data Assimilation and Optimization, provides modular data assimilation and optimization features in Python (\url{https://pypi.org/project/adao}) \cite{Argaud_2019}. It can be coupled with other modules or external codes while providing a number of standard and advanced data assimilation or optimization methods. The ADAO library also covers a wide variety of practical applications, from real engineering to experimental methodologies. Its architecture and numerical scalability adapt to the field of application. As shown by Eq. \ref{eq:post}, the optimal search for the control vector ${\bf{X}}$ takes a minimization form of an objective or cost function (given by the expression inside the exponential term) which must satisfy the background error statistics (prior term) and the equivalent observation error (likelihood term). This minimization process, equivalent to the maximum \textit{a posteriori} search, is carried out using the 3D-VAR algorithm. The control parameter is composed of the five most influential variables identified by the multivariate sensitivity analysis performed previously. The initial guess ${\bf{X}}_0$ is set to random values inside the constrained search space (${\bf{X}}_0=(K_{s1}=73.76,K_{s2}=83.62,K_{s3}=83.62,\text{ \-}\alpha = 0.9729, \gamma = 0.8611)'$). The observation vector ${\bf{Y}}$  is the free surface flow evolution extracted every $60$ seconds at the Verdon, Lamena, Pauillac, Fort Médoc, Bassens and Bordeaux locations from noon August $12$ to midnight August $14$, $2015$. The chosen optimization method involves computing the partial derivatives of the observation operator $G$ with respect to ${\bf{X}}$, a classical finite differences method with a differential increment set to $10^{-4}$.
The error background and observation covariance matrices respectively identified by $\mathcal{B}$ and $\mathcal{R}$ are token diagonals, meaning they have no error correlations. A small variance value for $\mathcal{R}$, justifying great confidence in the observation value, is considered, such as $\sigma_m^2=0.1*{\bf{Y}}$. On the contrary, little confidence is given to a prior part such as $\sigma_b^2=10.*{\bf{X}}_0$. As shown by Figures \ref{fig:cost_func}-\ref{fig:param_esti}, the automatic calibration algorithm finds an optimal solution in about $30$ iterations with the following set of parameters ${\bf{X}}_{MAP}=(K_{s1}=47.99,K_{s2}=59.63,K_{s3}=67.485,\alpha = 0.9114,\gamma=0.5344)'$. A rapid decrease in cost function can be observed for the first algorithm iterations.

The cost function curve behavior is smooth until iteration number $20$, where a slight decrease can be observed. As expected, a similar behavior is observed for the parameter to be calibrated.

\pagebreak

\begin{figure}[ht!]
\centering
\begin{tabular}{cc}
\subfloat[Verdon observation station\label{Verdon_calibrated}]{%
\includegraphics[width=0.5\textwidth]{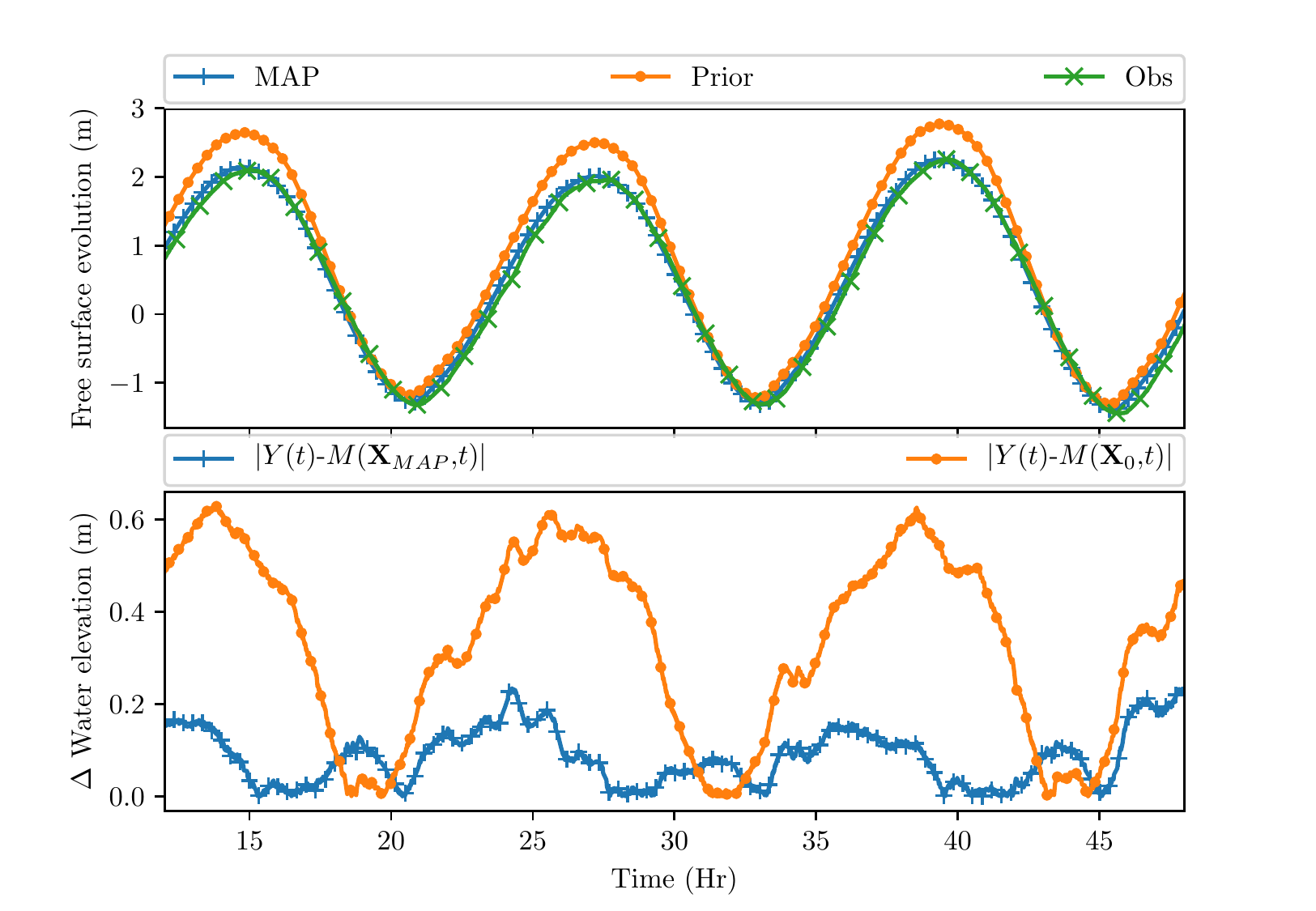}}&
\subfloat[Bordeaux observation station\label{Bordeaux_calibrated}]{%
\includegraphics[width=0.5\textwidth]{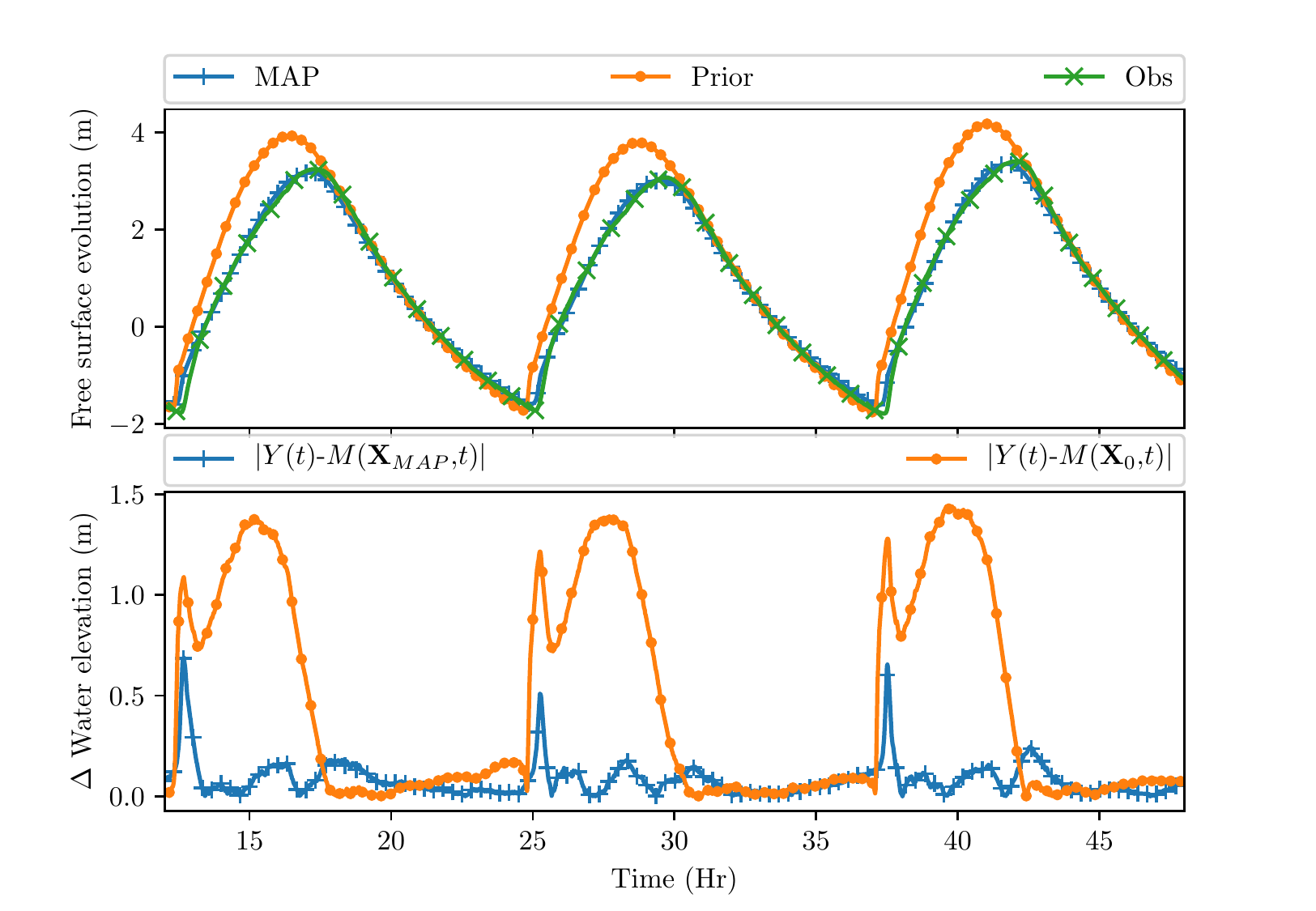}}
\end{tabular}
\caption{Comparison of the water depth evolution with and without calibration}
\label{fig:verdon_bordeaux_cali}
\end{figure}

Figure \ref{fig:verdon_bordeaux_cali} displays the results of automatic calibration over the computation period. As expected, the water surface profiles, calculated from the calibrated parameter configuration, are much closer to measurements than the ones computed from the background knowledge parameters. The final results emphasize the efficiency of the automatic calibration tool in the framework of a real configuration. For most of the studied period, the difference between observations and the calibrated configuration is less than $20$ $\text{cm}$ (about $5$\% of the tidal range) at the Verdon and Bordeaux stations. However, at the Bordeaux observation station, the calibrated configuration presents some error peaks at low tide. This phenomenon can be induced by a greater effect of the fluvial part of the estuary, which is not well represented in the model, where discharge is expressed as an hourly average.

\section{Discussion}
\label{sec:discussion}

The concept of interoperability is a generic solution for gathering and exchanging information from various multidisciplinary knowledge. The application, presented in this article, is a case of 2D hydrodynamics and is not representative of all the possibilities offered by the interoperability of \telemacsystem. For example, the estuarine sediment transport could be taken into account in order to better model and understand the evolution of the bed with the API of the \gaia module. More generally the approach presented here can be easily applied to different geoscience problems where \telemacmascaret is relevant. All APIs of the different modules are freely available as they are part of the \telemacsystem.

Applying interoperability criteria on an old and large code is not an easy task because of the transformation effort it requires. For the \telemacsystem the transformation towards the concept of interoperability required several years of development. A recommendation is to take into account these criteria at the early stage of development considering the fact that the code will probably have to interact with the outside world. The implementation of APIs makes it possible to extend the scope of the software by facilitating the use for different types of applications. For instance, based on fluid exchange of information, APIs can be used to couple the \telemacsystem with a Computational Fluid Dynamics (CFD)-type software (for example Code\_Saturne \citep{Archambeau_2004}; \url{www.code-saturne.org}) to take into account atmospheric and groundwater flows.

The development of standards for publishing interoperable softwares in forms suitable for community interactions remains a major issue \citep{Laniak_2013}. The present work allows the \telemacsystem to be integrated in different environments. However, the wrapper for each specific environment must be maintained. A lean standard has been proposed with the OpenMI environment \citep{Gregersen_2007}. The integration of the \telemacsystem in this environment could constitute an outlook to this work. Since Python is really easy to pick up and learn, a wrapper for this language is distributed in the \telemacsystem official version (TelAPy). This choice aims to favorise the dissemination of the hydro-informatic system as an environmental modeling tool. However, the system compilation is a cumbersome process which can slow down its dissemination. To overcome this issue, and based on the Python Package Index (PyPI, \url{https://pypi.org}), a compiled version of TelApy might be envisaged to provide all dynamic libraries needed to run \telemacsystem APIs.

To demonstrate that TelApy is a functional system, an example of hydraulic model calibration is presented. This case deals with a series of reference events by adjusting some uncertain physically based parameters until the comparison with observations achieves sufficient accuracy. If performed manually, the model calibration is time-consuming. Fortunately, the process can be largely automated to significantly reduce human workload, as shown in this paper. 

A reduced order model based on PCA-PCE is used to identify influential input parameters. This emulator has been created and validated on the basis of learning and validation solver computations. A major issue arising from this methodology concerns the optimization of the number of computational runs needed for sufficient results in the sensitivity analysis. Approximation with the surrogate models must become sufficiently accurate with just limited data available for the learning step. There have been some recent advances in this area, based on adaptive sampling \citep{Steiner_2019}. 

After most influential parameters have been identified, they are then calibrated using a data-driven technique. The chosen algorithm is based on a minimization process requiring derivatives of the \telemacdd solver with respect to the parameters to be calibrated. Several options exist to obtain the derivatives. The Finite Difference method, used in this work, is easily implemented but returns approximate derivatives whose poor accuracy can degrade the performance of the complete application. A much better option is to create a new program that computes the exact analytical derivatives of the model. Algorithmic Differentiation \citep{Griewank_2008} is a way of automating creation of the derivative program, thus providing accurate derivatives for a minimal development effort. Gradient-based methods are very useful to find local extreme values within a reasonable time, but they cannot pretend to find a global solution in the search domain. On the other hand, metaheuristic optimization algorithms are useful in finding the area of a global solution (minimum or maximum), but the convergence becomes much slower due to the large number of required simulations. One way to improve the efficiency of the calibration would be to combine a derivative-free algorithm (like a metaheuristic) with a gradient-based method (like BFGS) to obtain better solutions within a reasonable time (hybridization).

\section{Conclusion}
\label{sec:conclu}

The Application Programming Interface (API) of the open source \telemacsystem was developed to convert a heterogeneous set of open-source, user-contributed models into a suite of plug-and-play modeling components that can be reused in many different contexts. The APIs provide a user-friendly development framework that can be easily understood, allowing seamless integration of base codes, and do not invalidate existing institutional software development practices. The API development framework does not compel the use of, or supply, specific environmental modeling standards, as its services standardize at a more basic level of internal communication. This gives researchers and model developers more freedom to customize services for the problems they are facing.

The \telemacsystem API seeks to assist the hydrodynamic community in resolving new challenges, such as uncertainties, real-time data assimilation, and multi-physics simulations. Many model developers have limited skills in software development and architecture. Recognizing this and seeking to promote the widest possible adoption of the \telemacsystem API by the user community, a scripting feature of modeling is privileged with the Python language, as recommended by \citet{Knox_2018}. This forms a new module of the system called TelApy. Work performed with Python can be transposed to other scripting languages such as R and Julia. Python is a language that can be used in many contexts and adapted to any type of program based on dedicated libraries. However, it is particularly used to automate tedious tasks, such as the calibration process of a numerical model, saving a significant amount of time in realizing a project. Use of Python is widespread in the scientific community, and it has many libraries optimized and intended for numerical computation.

This flexibility was demonstrated in the Gironde Estuary case. First, a sensitivity analysis was carried out to identify sensitive parameters to calibrate. The most influential variables on water depth variation were the upstream friction coefficients ($K_{s1}$, $K_{s2}$, $K_{s3}$), the tidal amplitudes multiplier coefficient of tidal range $\alpha$, and the sea level correction $\gamma$. This was achieved by linking a \telemacdd physical-based component to an uncertainty quantification library. To achieve better model accuracy, a calibration process was realized through a physical based data-driven technique using a data assimilation library. To promote and facilitate the dissemination of the deployed approach to the \telemacmascaret community, the development of a Graphical User Interface (GUI) can be useful. The major benefits are user-friendliness, efficiency and enhancing the quality of hydraulic studies.

The calibration process deployed here can be extended to other solvers of the \telemacsystem (water quality, sediment transport, wave propagation and so on). Moreover, there are key potentially available sources of information on continental water bodies (in situ and remote sensing data, for instance). Data assimilation algorithms for integrating observation data into real cases are now increasingly applied to hydraulic problems with two main objectives: optimizing model parameters and improving stream-flow simulation and forecasting. Therefore, the ability to exchange data as computational results has become a growing need, facilitated by interoperability.

\section*{Acknowledgements}

The authors gratefully acknowledge contributions from the open-source community, especially that of OpenTURNS (Open source initiative for the Treatment of Uncertainties, Risks'N Statistics) and ADAO (a module for Data Assimilation and Optimization). In particular, we would like to thank Angelique Pon\c{c}ot and Jean-Philippe Argaud from EDF R\&D for their constructive discussion on data assimilation, Regis Lebrun from Airbus for the POD construction based on OpenTURNS and all of the EDF R\&D OpenTURNS team. The authors also would like to address special thanks to Kamal El Kadi Abderrezzak for his support and deep reading that greatly enriched this paper. Finally, we would like to thank the anonymous reviewer and Arnald Puy, whose comments and suggestions helped improve the manuscript.
\bibliographystyle{abbrvnat}
\bibliography{main}

\appendix
\setcounter{table}{0}
\renewcommand{\thetable}{\Alph{section}.\arabic{table}}

\section{\telemacdd{} list of API variables}
\label{appendix_apivar}
\lstset{rulecolor=\color{black}}
\begin{longtable}{|p{.30\textwidth} | p{.6\textwidth}|}
\hline
Variable name & Definition \tabularnewline
\hline
\hline
\text{\lstinline{MODEL.AT}} & Current time\tabularnewline
\text{\lstinline{MODEL.DT}} & Time step\tabularnewline
\text{\lstinline{MODEL.BCFILE}} & Boundary condition file name\tabularnewline
\text{\lstinline{MODEL.BND\_TIDE}} & Options for tidal boundary condition\tabularnewline
\text{\lstinline{MODEL.BOTTOMELEVATION}} & Bottom elevation\tabularnewline
\text{\lstinline{MODEL.CHESTR}} & Roughness coefficient\tabularnewline
\text{\lstinline{MODEL.FAIR}} & Wind influence coefficient\tabularnewline
\text{\lstinline{MODEL.COTE}} & Water level on boundary conditions\tabularnewline
\text{\lstinline{MODEL.CPL\_PERIOD}} & Coupling period with \telemacmascaret solvers\tabularnewline
\text{\lstinline{MODEL.DEBIT}} & Flowrate discharge on boundary conditions\tabularnewline
\text{\lstinline{MODEL.DEBUG}} & Activation debug mode keyword\tabularnewline
\text{\lstinline{MODEL.FLUX\_BOUNDARIES}} & Flux on the boundary conditions\tabularnewline
\text{\lstinline{MODEL.GEOMETRYFILE}} & Geometry file name\tabularnewline
\text{\lstinline{MODEL.METEOFILE}} & Binary atmospheric file name\tabularnewline
\text{\lstinline{MODEL.FO2FILE}} & Formatted data file name\tabularnewline
\text{\lstinline{MODEL.LIQBCFILE}} & Liquid boundary file name\tabularnewline
\text{\lstinline{MODEL.PREFILE}} & Previous computation file name\tabularnewline
\text{\lstinline{MODEL.GRAPH\_PERIOD}} & Graphical output period\tabularnewline
\text{\lstinline{MODEL.HBOR}} & Value of prescribed water depth on boundary conditions\tabularnewline
\text{\lstinline{MODEL.IKLE}} & Mesh connectivity table\tabularnewline
\text{\lstinline{MODEL.NACHB}} & Number of processors sharing a given node\tabularnewline
\text{\lstinline{MODEL.KNOLG}} & Pointer to get the global numbering of a node in the initial mesh\tabularnewline
\text{\lstinline{MODEL.KP1BOR}} & Boundary node connectivity\tabularnewline
\text{\lstinline{MODEL.LIHBOR}} & Boundary type for the water depth\tabularnewline
\text{\lstinline{MODEL.LISTIN\_PERIOD}} & Listing output period\tabularnewline
\text{\lstinline{MODEL.LIUBOR}} & Boundary type for horizontal velocity\tabularnewline
\text{\lstinline{MODEL.LIVBOR}} & Boundary type for vertical velocity\tabularnewline
\text{\lstinline{MODEL.L}\texttt{T}} & Current time step\tabularnewline
\text{\lstinline{MODEL.COMPLEO}} & Graphical output counter\tabularnewline
\text{\lstinline{MODEL.PTINIG}} & Number of the first graphical printout time step\tabularnewline
\text{\lstinline{MODEL.NPTIR}} & Number of border nodes for subdomains\tabularnewline
\text{\lstinline{MODEL.NBOR}} & Boundary node global number\tabularnewline
\text{\lstinline{MODEL.NELEM}} & Number of elements in the discretization mesh\tabularnewline
\text{\lstinline{MODEL.NELMAX}} & Maximum possible number of elements\tabularnewline
\text{\lstinline{MODEL.NPOIN}} & Number of nodes in the computational mesh\tabularnewline
\text{\lstinline{MODEL.NPTFR}} & Number of boundary nodes\tabularnewline
\text{\lstinline{MODEL.NTIMESTEPS}} & Number of time steps\tabularnewline
\text{\lstinline{MODEL.NUMLIQ}} & Liquid boundary number\tabularnewline
\text{\lstinline{MODEL.POROSITY}} & Porosity\tabularnewline
\text{\lstinline{MODEL.RESULTFILE}} & Result file name\tabularnewline
\text{\lstinline{MODEL.SEALEVEL}} & Sea level calibration coefficient\tabularnewline
\text{\lstinline{MODEL.TIDALRANGE}} & Tidal range calibration coefficient\tabularnewline
\text{\lstinline{MODEL.VELOCITYRANGE}} & Velocity range calibration coefficient\tabularnewline
\text{\lstinline{MODEL.UBOR}} & Value of prescribed horizontal velocity on boundary conditions\tabularnewline
\text{\lstinline{MODEL.VBOR}} & Value of prescribed vertical velocity on boundary conditions\tabularnewline
\text{\lstinline{MODEL.VELOCITYU}} & Horizontal velocity\tabularnewline
\text{\lstinline{MODEL.VELOCITYV}} & Vertical velocity\tabularnewline
\text{\lstinline{MODEL.WATERDEPTH}} & Water depth\tabularnewline
\text{\lstinline{MODEL.X}} & Horizontal coordinates of mesh points\tabularnewline
\text{\lstinline{MODEL.XNEBOR}} & Horizontal component of the normal at boundary nodes\tabularnewline
\text{\lstinline{MODEL.Y}} & Vertical coordinates of mesh nodes\tabularnewline
\text{\lstinline{MODEL.YNEBOR}} & Vertical component of the normal at boundary nodes\tabularnewline
\text{\lstinline{MODEL.EQUATION}} & Name of solved equation\tabularnewline
\text{\lstinline{MODEL.AK}} & Turbulent kinetic energy term of $k-\epsilon$ turbulence model\tabularnewline
\text{\lstinline{MODEL.EP}} & Dissipation term of $k-\epsilon$ turbulence model\tabularnewline
\text{\lstinline{MODEL.ITURB}} & Turbulence model\tabularnewline
\text{\lstinline{MODEL.INIT\_DEPTH}} & Initial value of water depth\tabularnewline
\text{\lstinline{MODEL.TRACER}} & Tracer value on mesh nodes\tabularnewline
\text{\lstinline{MODEL.NTRAC}} & Number of tracers\tabularnewline
\text{\lstinline{MODEL.FLOWRATEQ}} & Solid transport flowrate\tabularnewline
\text{\lstinline{MODEL.DCLA}} & Median grain size value\tabularnewline
\text{\lstinline{MODEL.SHIELDS}} & Critical Shields parameter\tabularnewline
\text{\lstinline{MODEL.XWC}} & Settling velocity\tabularnewline
\text{\lstinline{MODEL.Z}} & Free surface elevation\tabularnewline
\text{\lstinline{MODEL.QBOR}} & Value of prescribed discharge flowrate on boundary nodes\tabularnewline
\text{\lstinline{MODEL.EBOR}} & Value of prescribed turbulence dissipation term on boundary nodes\tabularnewline
\text{\lstinline{MODEL.FLBOR}} & Value of prescribed bottom elevation on boundary nodes\tabularnewline
\text{\lstinline{MODEL.TOB}} & Shear stress\tabularnewline
\text{\lstinline{MODEL.LIQBOR}} & Boundary type for the discharge\tabularnewline
\text{\lstinline{MODEL.NSICLA}} & Number of bed material size classes\tabularnewline
\text{\lstinline{MODEL.NOMBLAY}} & Number of layers in the bed\tabularnewline
\text{\lstinline{MODEL.CONCENTRATION}} & Concentration at the current time step\tabularnewline
\text{\lstinline{MODEL.EVOLUTION}} & Evolution of bed\tabularnewline
\text{\lstinline{MODEL.PARTHENIADES}} & Parthenaides erosion coefficient of each bed layer\tabularnewline
\text{\lstinline{MODEL.VOLU2D}} & Basis integral\tabularnewline
\hline
\caption{\telemacdd variables accessible through API}
\label{Tab:api_var}
\end{longtable}

\section{Programming the Gironde estuary case in Python}
\label{appendix_telapy}

\begin{lstlisting}[language=TelPython]
# Configuration step
#   'cas_file': telemac steering file
#   'for_file': User fortran folder
#   'comm': MPI communicator
t2d = Telemac2d(cas_file, user_fortran=for_file, comm=comm)

# Reading of TELEMAC-2D steering file and run memory allocation
t2d.set_case()

# Initialization of the TELEMAC-2D model
t2d.init_state_default()

# Six friction coefficient values are used
CHESTR = [CF1, CF2, CF3, CF4, CF5, CF6]
for k in range(len(CHESTR)):
    # Set the corresponding value to each node according to the area location
    for i in AREA[k]:
        t2d.set("MODEL.CHESTR", CHESTR[k], i=i)

# Set the tidal range value with the 'Hrange' variable
t2d.set("MODEL.TIDALRANGE", Hrange)
# Set the tidal velocity value with the 'Vrange' variable
t2d.set("MODEL.VELOCITYRANGE", Vrange)
# Set the sea level value with the 'SeaLevel' variable
t2d.set("MODEL.SEALEVEL", SeaLevel)

# MPI barrier
comm.Barrier()

# Initialize a list 'waterLevels' to store results in certain coordinate points
waterLevels = []
for j in xrange(len(self.observation_points)):
    waterLevels.append([])

# Time loop
ntimestep = self.t2d.get("MODEL.NTIMESTEPS")
for i in xrange(ntimestep):
    # Running one time step
    t2d.run_one_time_step()
    # Extract results if the time corresponds to observations
    if i >= debObs and i <= finObs and i % freqObs == 0:
        # Getting stations water level values
        for j in range(len(self.observation_points)):
            # 'H' is the water depth and 'Zf' the bathymetric level
            H, _, _, Zf = getStationValues(self.observation_points[j])
            waterLevels[j].append(H+Zf)

# Finalize
comm.Barrier()
t2d.finalize()
del self.t2d
\end{lstlisting}

\section{Bayesian inference formulation of calibration problem}
\label{appendix_bayesian}

The \textit{posterior} distribution $\pi\left({\bf{X}}|{\bf{Y}}\right)$ can be determined through the well-known Bayes rule (Eq. \ref{eq:bayes}).

 \begin{equation}
 \pi\left({\bf{X}}|{\bf{Y}}\right)=\frac{\pi\left({\bf{Y}}|{\bf{X}}\right)\pi\left({\bf{X}}\right)}{\int\pi\left({\bf{Y}}|{\bf{X}}\right)\pi\left({\bf{X}}\right)d{\bf{X}}}
 \label{eq:bayes}
 \end{equation}
 
The term $\pi\left({\bf{Y}}|{\bf{X}}\right)$, called the likelihood, can be interpreted as the probability density function of the observed data, conditional upon a set of parameter values (considered as random variables), and as a measure of the information provided by the observations on the parameter values. From Eq. \ref{eq:observation_space}, the likelihood is expressed as $\pi\left({\bf{Y}}|{\bf{X}}\right)\propto\exp(-\frac{1}{2}\left[{\bf{Y}}-G\left({\bf{X}}\right)\right]\mathcal{R}^{-1}\left[{\bf{Y}}-G\left({\bf{X}}\right)\right]')$. 

The term $\pi\left({\bf{X}}\right)$ represents \textit{a priori} knowledge of the unknown parameters ${\bf{X}}$. This term is classically taken as a multivariate normal distribution with known mean ${\bf{X}}_0$ (derived from measurement data or previous computation) and covariance matrix $\mathcal{B}\in\mathbb{R}^{p\times p}$ positive definite such as $\pi\left({\bf{X}}\right)\propto\exp\left(-\frac{1}{2}\left[{\bf{X}}-{\bf{X}}_0\right]\mathcal{B}^{-1}\left[{\bf{X}}-{\bf{X}}_0\right]'\right)$.

From the previous expressions of \textit{a priori} and likelihood, the \textit{posterior} distribution is given by the following equation:
 
\begin{equation}
\pi\left({\bf{X}}|{\bf{Y}}\right)\propto \exp\left(-\frac{1}{2}\left[{\bf{Y}}-G\left({\bf{X}}\right)\right]\mathcal{R}^{-1}\left[{\bf{Y}}-G\left({\bf{X}}\right)\right]'-\frac{1}{2}\left[{\bf{X}}-{\bf{X}}_0\right]\mathcal{B}^{-1}\left[{\bf{X}}-{\bf{X}}_0\right]'\right)
\label{eq:post}
\end{equation}

The maximum \textit{a posteriori} (MAP) is equivalent to the formulation of the optimal search of control vector ${\bf{X}}$, which must satisfy the \textit{a priori} error statistics $J_b = \frac{1}{2}\left[{\bf{X}}-{\bf{X}}_0\right]\mathcal{B}^{-1}\left[{\bf{X}}-{\bf{X}}_0\right]'$ and the equivalent observation error statistics $J_{obs} = \frac{1}{2}\left[{\bf{Y}}-G\left({\bf{X}}\right)\right]\mathcal{R}^{-1}\left[{\bf{Y}}-G\left({\bf{X}}\right)\right]'$.

\section{Generalized Sensitivity Analysis based on low discrepancy sequence of Sobol}

\begin{longtable}{p{.13\textwidth} |llll}
\hline
\multirow{2}{*}{\diagbox{Inputs}{\textbf{GSI}}} & \multicolumn{2}{c}{\textbf{Verdon}}&\multicolumn{2}{c}{\textbf{Bordeaux}}\\
& First Order & Total Order & First Order & Total Order\\
\hline
\hline
$K_{s1}$ &  $0.126$   & $0.133$  &$0.110$ & $0.156$\\
$K_{s2}$ &  $0.01$   & $0.0153$  &$0.358$ & $0.437$\\
$K_{s3}$ &  $2.29\times 10^{-5}$   & $7.45\times 10^{-5}$  &$0.0816$ & $0.118$\\
$K_{s4}$ &  $0.$   & $1.33\times 10^{-5}$  &$9.69\times 10^{-5}$ & $3.44\times 10^{-4}$\\
$K_{s5}$ &  $2.53\times 10^{-5}$   & $6.50\times 10^{-5}$  &$2.06\times 10^{-3}$ & $3.54\times 10^{-3}$\\
$K_{s6}$ &  $1.65\times 10^{-7}$   & $8.13\times 10^{-6}$  &$4.17\times 10^{-5}$ & $1.86\times 10^{-4}$\\
$\alpha$ &  $0.0464$   & $0.0492$  &$0.0135$ & $0.019$\\
$\beta$ &  $4.82\times 10^{-4}$   & $5.07\times 10^{-4}$  &$1.81\times 10^{-4}$ & $2.70\times 10^{-4}$\\
$\gamma$ &  $0.809$   & $0.809$  &$0.341$ & $0.364$\\
\hline
\hline
\caption{Generalized Sensitivity Indices (GSI) based on $1024$ $(2^{10})$ low discrepancy Sobol points at Bordeaux and Verdon observation stations}
\label{Tab:lds_GSI_bordeau}
\end{longtable}

\end{document}